\documentclass{siamonline181217}

\usepackage{amsmath,amsfonts,color,graphicx,algorithm,algpseudocode,dsfont,verbatim}

\definecolor{darkblue}{rgb}{0,0,0.9}

\renewcommand{\Delta}{\triangle}
 
\newcommand{\dee}{\mathrm{d}}

\newcommand{\sgn}{\mathrm{sgn}}

\newcommand{\R}{\mathbb{R}}
\newcommand{\N}{\mathbb{N}}
\newcommand{\cH}{\mathcal{H}}

\newcommand{\la}{\langle}
\newcommand{\ra}{\rangle}
\newcommand{\iid}{\overset{\mathrm{i.i.d.}}{\sim}}
\newcommand{\cL}{\mathcal{L}}
\newcommand{\cC}{\mathcal{C}}

\newcommand{\GP}{\mathsf{G}\mathsf{P}}

\newcommand{\sN}{\mathsf{N}}
\newcommand{\sU}{\mathsf{U}}

\newsiamremark{remark}{Remark}
\newsiamremark{hypothesis}{Hypothesis}
\crefname{hypothesis}{Hypothesis}{Hypotheses}
\newsiamthm{claim}{Claim}
\newsiamremark{example}{Example} %
\newsiamthm{assumptions}{Assumptions} %

\graphicspath{{./images/}}

\newcommand{\TheTitle}{Dimension-robust MCMC in Bayesian inverse problems} 
\newcommand{\TheAuthors}{V. Chen, M. M. Dunlop, O. Papaspilipoulos and A. M. Stuart}

\title{Dimension-Robust MCMC in Bayesian Inverse Problems}
\headers{Dimension-Robust MCMC Sampling}{\TheAuthors}

\author{
  Victor Chen\thanks{Computing \& Mathematical Sciences, California Institute of Technology, Pasadena, California, 91125, USA (\email{vlchen@caltech.edu}, \email{mdunlop@caltech.edu}, \email{astuart@caltech.edu).}}
  \and
  Matthew M. Dunlop\footnotemark[1]
  \and
  Omiros Papaspiliopoulos\thanks{ICREA and Department of Economics and Business Universitat Pompeu
Fabra, Barcelona 08005, Spain (\email{omiros.papaspiliopoulos@upf.edu})}
  \and
  Andrew M. Stuart\footnotemark[1]
}

\ifpdf
\hypersetup{
  pdftitle={\TheTitle},
  pdfauthor={\TheAuthors}
}
\fi

\begin{document}
\maketitle

\begin{abstract}
The methodology developed in this article is motivated by a  wide range of prediction and uncertainty quantification problems that arise in Statistics, Machine Learning and Applied Mathematics, such as non-parametric regression, multi-class classification and inversion of partial differential equations. One popular formulation of such problems is as Bayesian inverse problems, where a prior distribution is used to regularize inference on a high-dimensional latent state, typically a function or a field. It is common that such priors are non-Gaussian, for example piecewise-constant or heavy-tailed, and/or hierarchical, in the sense of involving a further set of low-dimensional parameters, which, for example, control the scale or smoothness of the latent state. In this formulation prediction and uncertainty quantification relies on efficient exploration of the posterior distribution of latent states and parameters. This article introduces a framework for efficient MCMC sampling in Bayesian inverse problems that capitalizes upon two fundamental ideas in MCMC, non-centred parameterisations of hierarchical models and  dimension-robust samplers for latent Gaussian processes. Using a range of diverse applications we showcase that the proposed framework is dimension-robust, that is, the efficiency of the MCMC sampling does not deteriorate as the dimension of the latent state gets higher. We showcase the full potential of the  machinery we develop in the article in semi-supervised multi-class classification, where our sampling algorithm is used within an active learning framework to guide the selection of input data to manually label in order to achieve high predictive accuracy with a minimal number of labelled data. 
  
\end{abstract}

\section{Introduction}
\label{sec:intro}

\subsection{Overview}
\label{ssec:bg}

The methodology developed in this article is motivated by a  wide range of learning problems that arise in Statistics, Machine Learning and Applied Mathematics. We illustrate our methods on non-parametric regression, inference for functions involved in differential equation models and semi-supervised classification; \Cref{ssec:mote} provides a quick overview of those.  
Despite the significant and apparent differences in their details all these learning problems can be formulated as ill-posed inverse problems. 
There is data  $y$ assumed to arise from a (forward) model $G$ via
\begin{align}
\label{eq:data}
y_j = G(u;\eta_j)\,,\quad y = \{y_j\} 
\end{align}
where $\eta_j$ are i.i.d. realisations of some random noise, and we wish to
recover the latent function  $u(x)$ for $x \in D$, where typically $D \subseteq \R^d$. This is ill-posed in the sense that $y$ contains little or no signal, relative to noise, for parts 
of  $u$, and some type of
regularisation is needed to obtain a useful inversion. One canonical way to regularize and also deal with uncertainty quantification is to use Bayesian modelling and assign a prior distribution on $u$, $\mu_0(\dee u | \theta)$.
The data model \cref{eq:data} leads to the likelihood
\[
\cL(y|u) = \exp(-\Phi(u;y))
\]
of $y$ given $u$, which is combined with $\mu_0$ using Bayes' theorem, to produce the posterior distribution $\mu^y$:
\[
\mu^y(\dee u|\theta) \propto \exp(-\Phi(u;y))\mu_0(\dee u |\theta).
\]
The prior typically  involves a low-dimensional parameter vector $\theta$, the choice of which is critical for good performance (e.g., reconstruction of unknown function, prediction of future data, etc.) and it is best done using a data-driven criterion. One approach is to adopt priors for the parameters, $\pi_0(\theta) \dee \theta$  and carry out Bayesian inference for the latent state and parameters, on the basis of the posterior $\mu^y (\dee u, \dee \theta)$. This article proposes a simple, efficient and plug-and-play methodology for sampling such posteriors using MCMC. To fix some terminology, we will call the models described above as latent Gaussian when $\mu_0(\dee u|\theta)$ defines a Gaussian distribution, and as latent non-Gaussian otherwise, and as hierarchical when $\pi_0(\dee \theta) \dee \theta$ is part of the model. The article is interested in latent non-Gaussian and/or hierarchical models. 

The article builds upon two fundamental approaches for efficient sampling in such frameworks both of which have appeared in Statistical Science. \cite{papaspiliopoulos2007general} described a generic framework for efficient sampling of latent states and parameters in Bayesian hierarchical models, the so-called non-centred algorithms. These are based on identifying a transformation $u = T(\xi,\theta)$ such that $\xi$ and $\theta$ are a priori independent, and then sampling from the transformed posterior $\nu^y (\dee \xi, \dee \theta)$ by iterative sampling from the two conditionals, $\nu^y (\dee \xi | \theta)$ and $\nu^y (\dee \theta | \xi)$; samples can be transformed then  to $u=T(\xi,\theta)$. This scheme has been found to be extremely successful in a large number of contexts, but it requires that $\nu^y (\dee \xi | \theta)$ can be sampled efficiently. The latent states are high-dimensional, and in certain formulations infinite-dimensional, hence their conditional posterior sampling is challenging.
\cite{cotter2013mcmc} systematized and popularized in the Statistics community an MCMC framework for efficient sampling of high-dimensional latent states in latent Gaussian models.  The so-called preconditioned Crank-Nicolson (pCN) algorithm (see \Cref{ssec:rob2} later in this article for a description) samples from $\mu^y(\dee u | \theta)$ when $\mu_0(\dee u)$ is Gaussian and is well-defined even when $u$ is infinite-dimensional, and correspondingly $\mu_0$ is a Gaussian process prior. This has then the effect that when $u$ (correspondingly,  $\mu_0$ and the forward problem) has been discretized to finite dimension $N$, the performance of pCN that samples from the discretized posterior is robust to the choice of $N$. In simple terms, the number of samples needed to
approximate a given posterior quantity at a specified accuracy does 
not grow as the discretization gets finer, i.e., as $N$ gets larger; we refer to an algorithm that enjoys this property 
as {\em dimension-robust}.  Practically,
Metropolis-Hastings algorithms failing to meet this criterion 
manifest themselves as having to scale the proposed step by smaller and smaller
constants, as the dimension of the target gets larger, in order to maintain 
the same acceptance probability. 
It is known that
na\"ive application of Metropolis-Hastings algorithms in Bayesian inverse problems
will suffer from the deterioration described above.  A range of other dimension-robust but also gradient-based Metropolis-Hastings algorithms for latent Gaussian models has appeared since  \cite{cotter2013mcmc}, for instance
\cite{BGLFS17,cui2016dimension,law2014proposals,titsias2016auxiliary}; apart from pCN in this article we also explicitly consider the  $\infty$-MALA 
and the $\infty$-HMC algorithms from this new generation. 

The methodology we develop in this article puts in a single framework dimension-robust sampling of latent Gaussian states and non-centred parameterizations of hierarchical models to produce a simple, efficient and plug-and-play framework for posterior sampling in Bayesian inverse problems that are non-Gaussian and/or hierarchical. The idea is to identify the prior-orthogonalizing transformation $u=T(\xi,\theta)$ such that $\xi$ is a white noise process. Think of a white noise process as an infinite vector of i.i.d. standard Gaussians, $\xi_j\iid \sN(0,1)$. We can use a dimension-robust sampler for $\nu^y(\dee \xi | \theta)$, such as pCN or $\infty$-HMC, by exploiting that this is a latent Gaussian posterior distribution. When the parameters $\theta$ are part of the inference, we can use a non-centred algorithm, where now we exploit that transformed latent states can be sampled efficiently. Hence, the non-centred transformation to white noise solves simultaneously the two challenging sampling problems. Additionally, we unify two disparate themes of
current high interest in Bayesian inference: the use of transformations
from non-Gaussians to Gaussians as introduced within the randomise-then-optimise methodology in \cite{marzouk}; and the use of
non-centred parameterisations for hierarchical sampling.

Our article shows how to obtain the white noise representation of Gaussian processes and of a range of commonly used non-Gaussian processes and  provides dimension-robust algorithms for sampling in latent non-Gaussian and hierarchical inverse problems.  A main contribution of the article is the illustration that the proposed methodology provides excellent numerical results in a range of examples.

\begin{figure}
\begin{center}
\includegraphics[scale=0.4]{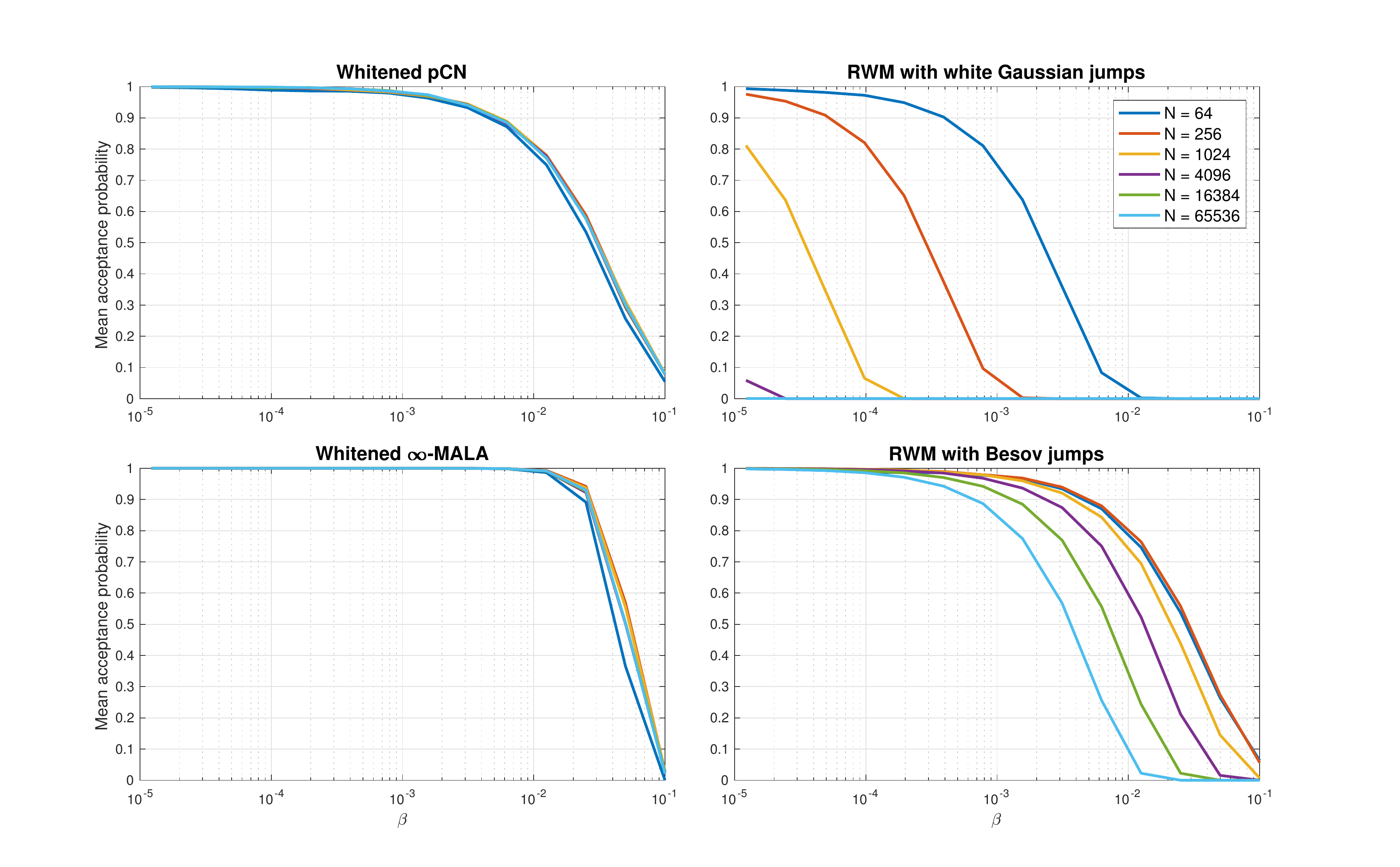}
\end{center}
\caption{Results for non-Gaussian priors based on series expansions: in this example a sparsity-inducing Besov prior is used within a regression example, the model is introduced as \Cref{ex:reg} in \Cref{ssec:mote}. The plot shows expected acceptance probability versus jump size $\beta$ for the whitened pCN algorithm, the whitened $\infty$-MALA algorithm and two random walk Metropolis algorithms. Curves are shown for different discretization sizes $N$ (the higher $N$ the finer the discretization). Details of the different algorithms can be found in \Cref{sub:sim-ng}.}
\label{fig:pcn_vs_rwm}
\end{figure}

\Cref{fig:pcn_vs_rwm} demonstrates what can
be achieved by applying the methodology herein.  The results correspond to a regression analysis (see \Cref{ex:reg} in \Cref{ssec:mote}), where the latent state is the regression function a priori modelled using a
Besov prior. Model-wise we have an analogue of Gaussian process regression but where the  infinite-dimensional analogue of a Bayesian lasso is used to model the regression function instead of a Gaussian process, see \Cref{sub:sim-ng} for details.
The figure illustrates the deterioration of two variants on the 
random walk Metropolis (RWM) algorithm, together with the dimension-robustness of the proposed 
whitened pCN  (wpCN) and the whitened $\infty$-MALA 
 (w$\infty$-MALA) algorithms. In this example parameters are kept fixed and only latent states are sampled. The figure shows the acceptance probability as a function of
$\beta$, which controls the proposed jump size, as the discretization of 
the unknown (of the order $1/N$) becomes finer. 
For the proposed algorithms shown on the left column 
we obtain a dimension-independent scaling whereas RWM algorithms shown on the right column deteriorate with increasingly fine
discretisation levels.

\begin{figure}
\begin{center}
  \includegraphics[width=\textwidth,trim = 4cm 0cm 1cm 0cm, clip]{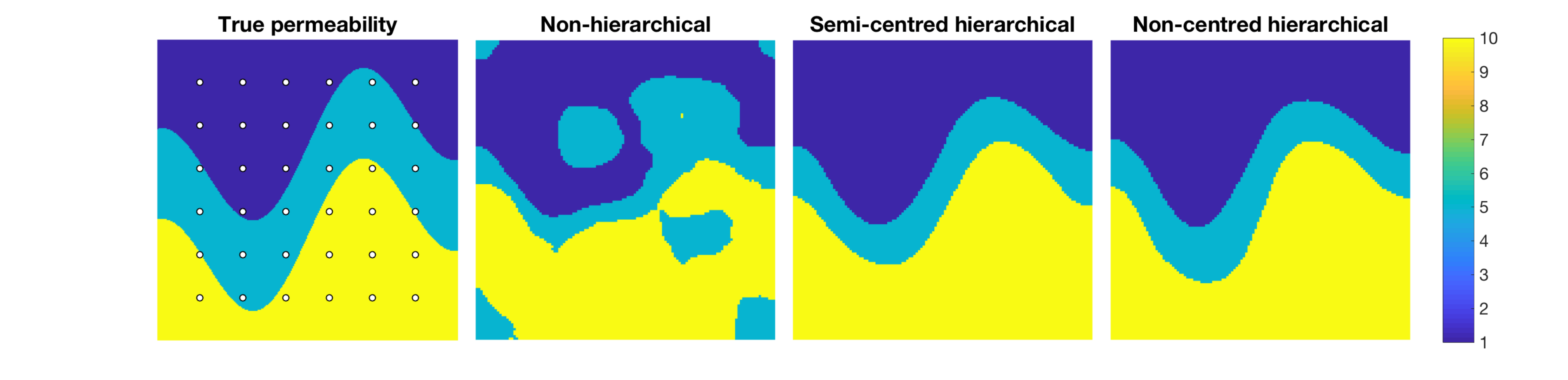}\\
  \includegraphics[width=0.64\textwidth]{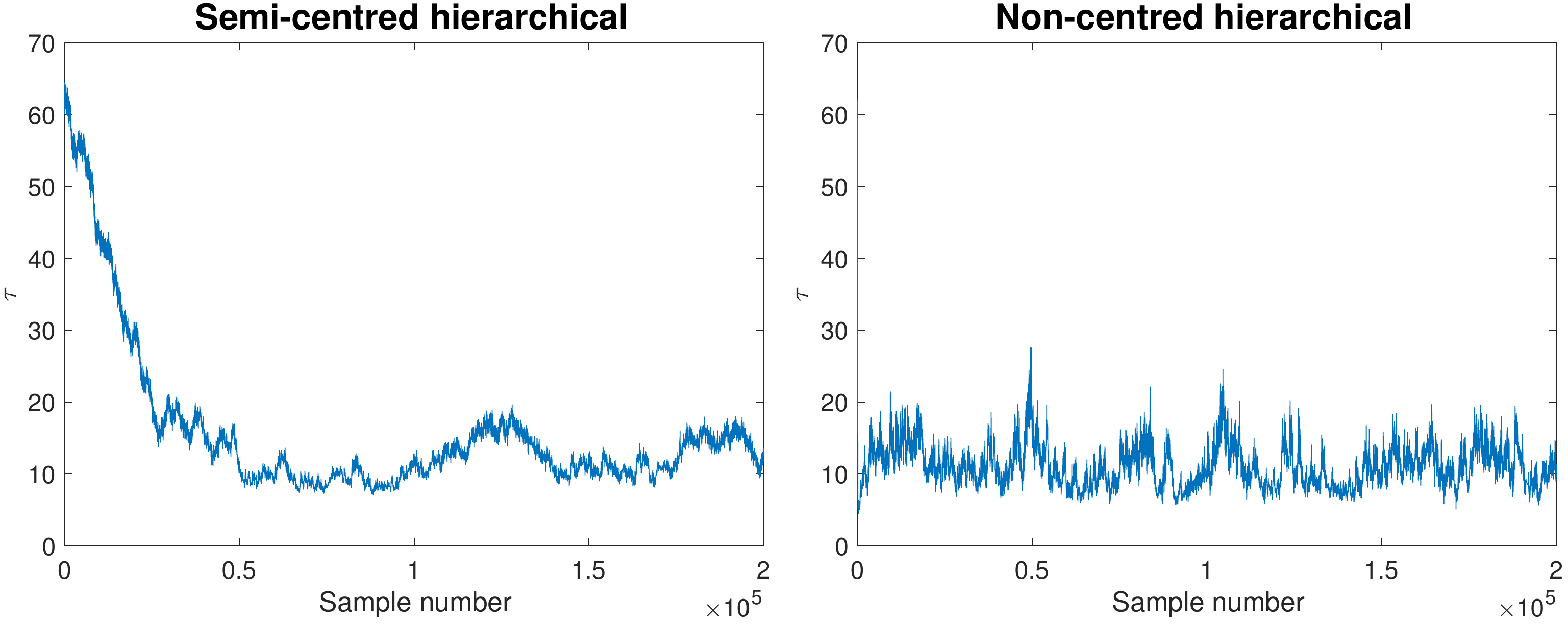}
\includegraphics[width=0.323\textwidth, trim = 0cm -0.4cm 0cm 0cm]{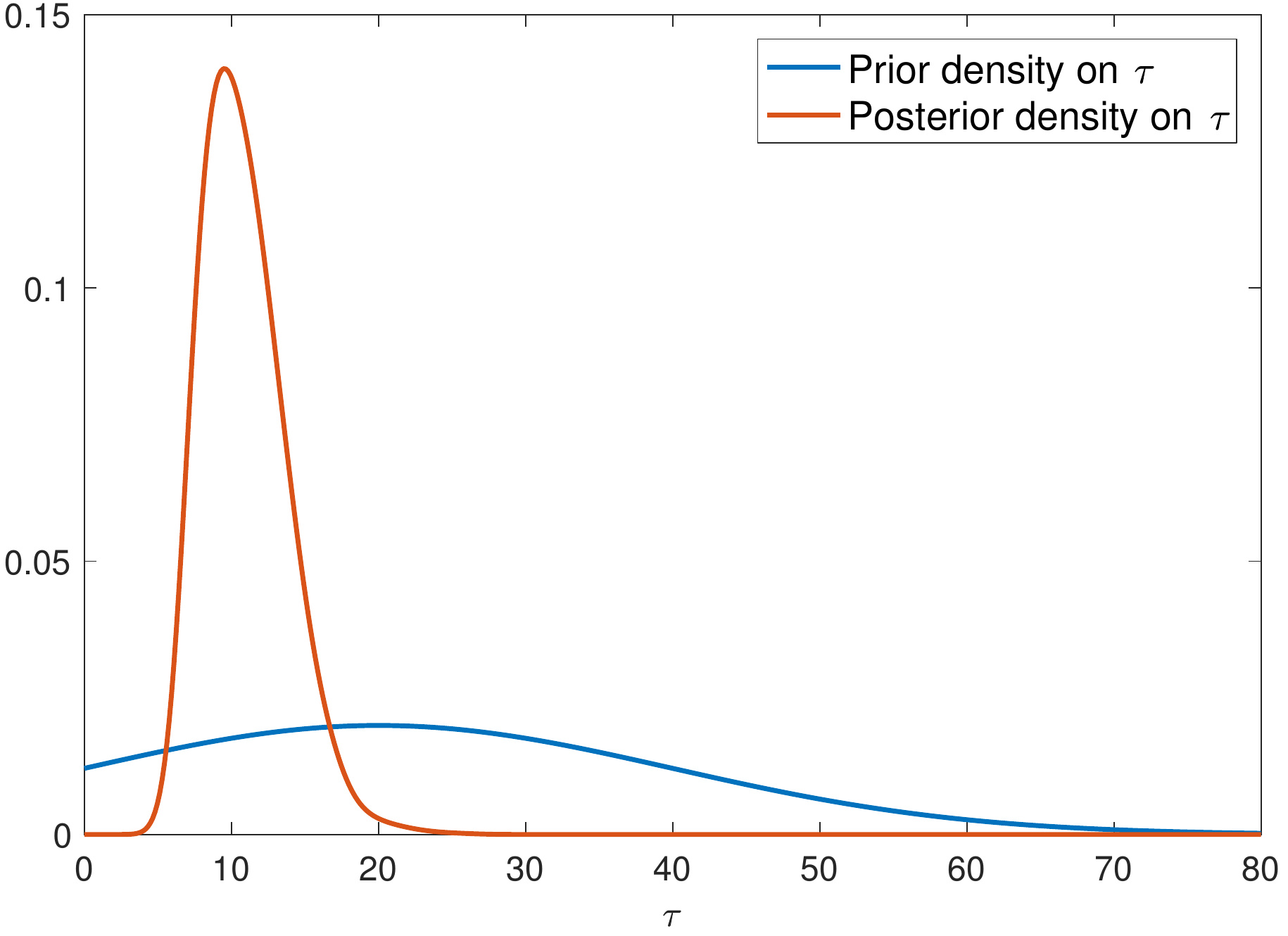}
\end{center}
\caption{Results for hierarchical priors: a hierarchical model with Gaussian latent state and prior on the length-scale parameter that defines its covariance kernel in the context of recovering a latent permeability field on the basis of noisy pressure measurements; the model is described as \Cref{ex:inv_pde} in \Cref{ssec:mote}. Top: The log-permeability that we wish to recover and the observation locations (left) and the conditional means from the three methods considered - the proposed is the rightmost. Details in \Cref{ssec:darcy}. Bottom: The trace of the length-scale parameter when using the semi-centred hierarchical method (left) versus the proposed non-centred method (middle). The parameter takes approximately 250 times longer to burn-in with the semi-centred method, whilst also exhibiting poorer mixing. A comparison of the prior and posterior density on the length-scale parameter $\tau$ for the non-centred hierarchical model is also shown (right). }
\label{fig:gwf_truth}
\end{figure}

\Cref{fig:gwf_truth}(top) shows the recovery of a permeability field on the basis of a small number of indirect and noisy observations of pressure on the input locations shown in the Figure, where the permeability and pressure are related though a partial differential equation model, see \Cref{ex:inv_pde} in \Cref{ssec:mote} for a brief description and  \Cref{ssec:darcy} for details. The proposed method corresponds to the right-most reconstruction. The quality of the reconstruction relies of exploring successfully the posterior distribution of a parameter that controls the smoothness of the field - this can be appreciated by comparing the reconstruction on the second panel that does not learn this parameter with the other two that do.  Therefore, this example highlights the potential of our approach for hierarchical priors. The MCMC traces for the smoothness parameter are shown in \Cref{fig:gwf_truth}(bottom). The proposed method is the one with better mixing. An additional advantage of the proposed method is that it is plug-and-play, it relies very marginally on the details of the model. The other method that does reasonably well here - termed semi-centred in the Figures - is tailored to this specific model.  


\section{Problem and  models}
\label{ssec:mote}

We give three  
examples that will be used throughout the paper  to showcase the performance of the methodology 
we advocate in this article. We also give an overview of the different prior models we employ for the latent states. 

\begin{example}[Regression]
  \label{ex:reg}
  Regression is the canonical example of an infinite dimensional inference
problem in Statistics. The goal is recovery of function $u(x)$ defined on a bounded
set $D \subset \R^d$ from a set of noisy  point evaluations $y_j = u(x_j) + \eta_j, j \in Z$. The formulation of regression as an inverse problem was sytematized in the
foundational work of  \cite{wahba1990spline}; subsequently Gaussian process 
based learning has built on this work in numerous directions \cite{williams2006gaussian}.
A more general version of this problem is to consider recovery of function $u$
from noisy observations of the action of a linear operator $K$ on $u$; in particular
it is particularly challenging to undertake this task when $K$ is compact meaning
that inversion is unstable, for example when $K$ is a convolution kernel such as
Gaussian blurring. The data may be written in the form  $y_j = (Ku)(x_j) + \eta_j$.  
We consider such an example in \Cref{sub:sim-ng}.  
More generally both of these problems may be cast in the general setting of
recovering function $u$ from $|Z|$ noisily observed linear functionals.
Posterior consistency for such problems is discussed in \cite{knapik2016bayes,agapiou2013posterior,stuart2018posterior}.

The results shown in \Cref{fig:pcn_vs_rwm} correspond to an analysis of 
the basic linear regression problem with a non-Gaussian prior. This prior
may be viewed as a functional version of Bayesian lasso in that the MAP estimator
minimizes the least squares funtional with lasso regularization. The prior is expressed as a series expansion,
$ u(x) = \sum_{j=1}^\infty \rho_j\zeta_j\varphi_j(x)$,
where $\rho = \{\rho_j\}_{j\geq 1}$ is a deterministic real-valued sequence, $\zeta_j$ are independent scalar random variables, $\zeta_j \sim G_j(\dee \zeta)$, and $\{\varphi_j(x)\}_{j\geq 1}$ are basis functions defined on $D$. This construction is inspired by an analogous one for Gaussian processes, the so-called Karhunen-Loeve expansion, which we recall in \Cref{ssec:KL}.  \Cref{ssec:white3} provides several specific instances of non-Gaussian expansions, motivations for their use in applications and links to the literature, together with their white noise representations. The prior behind the results in \Cref{fig:pcn_vs_rwm} corresponds to Laplace-distributed coefficients $\zeta_j$;  details on the specific prior, discretizations, and MCMC settings are provided in \Cref{sub:sim-ng}.  
\hfill\qed
\end{example}

\begin{example}[Graph-based semi-supervised Classification]
  \label{ex:inv_class} This problem concerns the prediction of labels for a large number of unlabelled input data, $x_j$, $j \in Z$, $x_j \in D \subset \R^d$, on the basis of a much smaller number of labelled input data, i.e., pairs of input and output observations $(x_j,y_j)$, $j \in Z' \subset Z$, where the output is categorical and without loss of generality taking values $y_j \in  \{1,\ldots,k\}$. \Cref{sec:MNIST} develops models and algorithms for this problem and illustrates them on the well-known MNIST dataset in which the labels are digits from 0 to 9 and the input are greyscale pixel intensities of hand-written digits. The approach we pursue is probabilistic classification, which explicitly accounts for misclassification errors. To every input data point $x_j$ we assign  latent variables $v_{j,r}$, $r=1,\ldots,k$, where for a given $j$,  each $v_{j,r}$ contributes to the probability of assigning the data point to the $r$th class. Let $v_{\cdot,r} = \{v_{j,r}\}_{j \in Z}$ and $v = \{u_{\cdot,r}\}_{r=1}^k$. We treat the $v_{\cdot,r}$'s a priori as i.i.d., assigned an $N$-dimensional Gaussian prior distribution, where $N=|Z|$, whose covariance matrix is constructed from the input data $\{x_j\}_{j \in Z}$. In particular, we identify the input data with the nodes of a graph and use  ideas from spectral clustering to identify connected components in the graph and use this spectral information to define the eigenstructure of the prior covariance. The prior covariance relies on two parameters that we treat hierarchically. The latent variables $v$ are mapped to latent variables $u$, which are one-hot-encodings of the classes, hence we define what we call in \Cref{sssec:lvl_prior} a vector level-set prior for $u$. We are interested in classifying a large number 
of inputs, i.e., taking $|Z|=N \gg 1$, on the basis of a small number of labelled inputs, i.e., $|Z'| \ll N$. In \Cref{sec:MNIST} we also implement an active learning approach, whereby on the basis of a very small number of labelled inputs and subsequent analysis with our model, we identify the inputs hardest to classify, obtain the labels of those 
(a human-in-the-loop approach) and re-run our analysis; this  demonstrates excellent predictive performance. The dimension-robustness of MCMC is critical here since $N$ is in the tens of thousands. Recent theory demonstrates that the discrete-space model we employ here for regularization has a continuum limit \cite{dunlop2018large}, and the $N$-dimensional Gaussian converges to a Gaussian process; we conjecture that the existence of the 
continuum limit  explains the empirically observed dimension-robustness of our sampler 
respect to dimension.
Using the connection between Whittle-Matern processes and samples from (possibly
fractional) partial differential equations, as established in \cite{matern_spde},
this limit may be viewed as a generalized Whittle-Mattern process; it is generalized
in the sense that the homogeneous Laplacian is replaced by an inhomogeneous
elliptic differential operator with coefficients depending on the sampling
density of the points $\{x_j\}.$ This connection allows for the transfer of
ideas from hierarchical sampling with Whittle-Matern processes into this graph
based setting; in particular the graph Laplacian plays the role of the Laplacian
in the methdology developed in \cite{matern_spde}. 
\hfill\qed 
\end{example}

\begin{example}[PDE-based Inversion]
  \label{ex:inv_pde}
These types of problems arise in numerous applications such as those
referenced in the lecture notes \cite{lecturenotes}. From a statistical
point of view their key feature is that the data is typically given as
linear functionals of a function (the output) which is itself linked to the 
unknown (the input) via a complex nonlinear transformation. This nonlinear transformation
is defined through solution of a partial differential equation, and
may involve expensive computer code. Indeed the celebrated work of
Kennedy and O'Hagan was concerned with methodologies which replace
the expensive computer code by a Gaussian process emulator \cite{ko01}.
Our work, however, takes a different path. Acknowledging that the
cost of the computer code evaluation will grow with the degree
of spatial resolution used in it, we seek to design MCMC methods whose
rate of convergence is independent of this level of spatial resolution.

The mapping from the input to the output is refered to as the forward model;
it is this map which is emulated in the approach \cite{ko01}.
The inverse problem refers to recovering the input from noisily-observed
linear functionals of the output; we thus view the input as a latent state and model it
with a prior probability distribution, in order to find the posterior given data.

We consider two specific examples of such nonlinear PDE-based inverse
problems in \Cref{ssec:darcy}; one is drawn from medical imaging and 
the other from fluid dynamics. For illustrative purposes we summarise 
the fluid mechanics problem below; results from it are shown in 
\Cref{fig:gwf_truth}.  The  steady state Darcy equation describes the 
pressure $p(x), x \in D$, where  $D\subset \R^2$ (one can define this for general $\R^d$ but this is beyond the scope of this example) in a porous medium. The pressure
solves the partial differential equation (PDE) 
\[
\begin{cases}
\hfill -\nabla\cdot(u\nabla p) = f & x \in D\\
\hfill p = 0 & x \in \partial D
\end{cases}
\]
where $f(x)$ denotes the sources and sinks of fluid,
$\nabla$ denotes the gradient vector and $\nabla \cdot$ its contraction to
a divergence, and $\partial D$ the boundary of $D$; think typically of $\partial D$ as 
including part of the ground surface and $x \in D$ a location underground).  The data are given by noisy measurements of pressure on different locations, $y_j = p(x_j) + \eta_j$. 
The latent unknown input state here is the spatial function $u(x), x \in D$, which determines the permeability of the
porous medium. 

A commonly occurring prior model is that the permeability is piecewise constant, with unknown interfaces between known constant values, for example corresponding to different types of subsurface rocks.  A latent non-Gaussian model is necessary for such piecewise constant latent function. We construct such a model by the so-called level-set method, which thresholds at different levels an underlying smooth Gaussian field. We use the Whittle-Matern Gaussian process as a model for the smooth field, which we recall in \Cref{ex:whittlematern} and give its white noise representation. The thresholding of the smooth Gaussian to obtain a piecewise constant prior is closely connected to the probit model in Statistics and to the approach we undertake in the graph-based semi-supervised learning in \Cref{ex:inv_class}, and \Cref{sssec:lvl_prior}  and \Cref{sec:MNIST} clarify these connections. The prior depends on parameters that we treat hierarchically, and \Cref{fig:gwf_truth} shows reconstruction results based on integrating out the
hierarchical parameters, as well as MCMC trace plots for one of the hierarchical parameters. 
\hfill\qed
\end{example}

\section{Good algorithms, bad theory (and vice versa)} 
\label{sec:step_choice}

The methods and experiments in the article involve local Metropolis-Hastings algorithms, such as random walk Metropolis, pCN, $\infty$-MALA etc, for sampling high-dimensional latent states.  All these algorithms involve the choice of a step size parameter that is denoted by $\beta$ throughout the article, see e.g. \Cref{alg:pcn} in \Cref{ssec:rob2}. The claim in the article is that the proposed algorithms   are dimension-robust, i.e.,  $\beta$ does not have to be a decreasing function of the dimension of latent states in order to  obtain comparable performance as that dimension varies, e.g. comparable acceptance probabilities.  Effectively, the combination of inverse problem and sampling algorithm makes the sampling problem a low-dimensional one. This concept is closely related to the definition of intrinsic dimension for importance sampling algorithms for linear inverse problems in \cite{importance_sampling}.

This claim is only supported via numerical experiments in a wide range of settings and for a wide range of applications. From a practitioner point of view there are two sets of theoretical results that would be useful to have. The first is a set of realistic conditions under which the claimed dimension-robustness is proven to hold. The second concern the optimal choice of $\beta$. We have not been able thus far to prove dimension-robustness under realistic weak assumptions, but we expect to do so in future work.  However, it is important to understand that optimal scaling theory for dimension-robust algorithms is inherently intangible.  This is a manifestation of what we call a ``good algorithms - bad theory'' situation. The fact that the combination of inference problem and sampling algorithm makes the sampling problem effectively low-dimensional, makes it  practically impossible to come up with a generic optimality theory for Metropolis-Hastings algorithms. Effectively every sampling problem is like a different low-dimensional problem, hence a good step size depends on the details of the model and the data and will be found by familiar pilot tuning (e.g. monitoring autocorrelations, etc.) or by using some version of adaptation. Good optimality theory exists for bad algorithms in high-dimensional regimes! The vast literature on optimal scaling, e.g., as in \cite{roberts2001optimal,beskos2013optimal}, applies to dimension-sensitive algorithms (more precisely, combinations of inference problems and sampling algorithms), e.g., choose $\beta$ so that the acceptance probability is 0.234.  Even for general Gaussian Bayesian inverse problems it is not possible to develop such optimal scaling theory for dimension-robust algorithms. (There is a common misconception that \cite{cotter2013mcmc} promote the rule of choosing $\beta$ in pCN so that to achieve an acceptance probability around 0.3 - this just happened to work for some of their experiments but it is not actually suggested as a principle, and there are no good reasons for this choice beyond those experiments!).    The main guideline is to choose $\beta$ so that the acceptance probability is not too small nor too large. The experiments in this article were tuned following this very weak criterion.   

\section{Gaussian priors} 
\label{sec:transformations}

This section first recalls a dimension-robust sampler for latent Gaussian models, the  pCN \Cref{alg:pcn}. The section provides a range of white noise representations of Gaussian processes. These have direct application to defining dimension-robust algorithms for hierarchical priors in \Cref{sec:hierarchical}, but are alaso indirectly useful for white noise representations of non-Gaussian processes. The Section also introduces, and studies from a variety of angles, the  Whittle-Mat\'ern Gaussian process that is used as a building block in several of our examples. The reader might find useful to consult \Cref{ex:whittlematern} while going through the white noise representations in \Cref{ssec:white2}, in order to have a concrete example to consider.

\subsection{Robust algorithms}
\label{ssec:rob2}

The 
preconditioned Crank-Nicolson (pCN) for Bayesian inverse problems applies when the
prior $\mu_0$ is the centred Gaussian $\sN(0,\cC)$ and it is presented in 
\Cref{alg:pcn}. Notice that the algorithm makes sense when $\sN(0,\cC)$ is both finite-dimensional and infinite-dimensional. The algorithm as named
is highlighted in the review paper \cite{cotter2013mcmc}, but is due to
Alex Beskos who, in \cite{beskos2008mcmc}, recognized it as a derivative-free 
simplification of the MALA-based dimension robust samplers introduced in
\cite{stuart2004conditional}. 

\begin{algorithm}
\caption{Preconditioned Crank-Nicolson (pCN)}
\label{alg:pcn}
\begin{algorithmic}[1]
\State Fix $\beta \in (0,1]$. Choose initial state $u^{(0)}\in X$.
\For{$k=0,\ldots,K-1$}
\State  Propose $\hat{u}^{(k)} = (1-\beta^2)^{\frac{1}{2}}u^{(k)} + \beta\zeta^{(k)},\quad\zeta^{(k)} \sim \sN(0,\cC)$.
\State \parbox[t]{\dimexpr\linewidth-\algorithmicindent}{Set $u^{(k+1)} = \hat{u}^{(k)}$ with probability
\[
\min\Big\{1,\exp\big(\Phi(u^{(k)};y) - \Phi(\hat{u}^{(k)};y)\big)\Big\}
\]
or else set $u^{(k+1)} = u^{(k)}$.\strut}
\EndFor
\State\Return $\{u^{(k)}\}_{k=0}^{K}$.
\end{algorithmic}
\end{algorithm}

\subsection{White noise representation of Gaussian priors}
\label{ssec:white2}

We now construct white noise representations of Gaussian processes. 
To be concrete we consider the case of Gaussian prior probability measure 
$\mu_0$ on a separable Banach space $X$ of continuous real-valued functions 
defined on bounded open $D\subseteq\R^d$. Measure $\mu_0$ can then be 
characterised by its mean function $m:D\to\R$ and covariance kernel 
$c:D\times D\to \R$, and is then commonly written as 
\[
\mu_0 = \GP(m(x),c(x,x')).
\]
The covariance function can be used to define a symmetric positive 
semi-definite covariance operator $\cC:X\to X$ by
\begin{align}
\label{eq:cov_op_kernel}
(\cC\varphi)(x) = \int_D c(x,x')\varphi(x')\,\dee x'
\end{align}
for any $\varphi \in X$, $x \in D$; we can then alternatively write
$\mu_0 = \sN(m,\cC)$. The operator $\cC$ is  
trace class, which implies its representation by a countable number of eigenvalues and eigenfunctions, in a way analogous to the spectral decomposition of covariance matrices, see \Cref{ssec:KL}. The inverse of the covariance operator, the so-called precision operator $\cL:D(\cL) \to X$, 
has domain $D(\cL)$ which is dense in $X$. For the definition of white noise as a Gaussian process on a Hilbert space see the Appendix.  
We now give three examples of white noise representations
of $\mu_0.$ We write $u= T(\xi)$ instead of $u = T(\xi,\theta)$; the dependence of the transformation on parameters $\theta$ is crucial in hierarchical models and we are explicit when dealing with  those  in \Cref{sec:hierarchical}.

\subsubsection{Cholesky factorisation of covariance matrix.}
\label{ssec:chol}
Consider the centred Gaussian
process $\mu_0 = \GP(0,c(x,x'))$ restricted to a set of $n$ points 
$D_n=\{x_j\}_{j=1}^n \subset D$.  Define matrix $C_n \in \R^{n\times n}$ by 
$(C_n)_{ij} = c(x_i,x_j)$ and denote by $C_n = Q_nQ_n^*$ its Cholesky 
decomposition. If 
\[
\Xi = \R^n,\quad \nu_0 = \sN(0,I),\quad T(\xi) = Q_n\xi
\]
then $(\Xi,\nu_0,T)$ is a white noise representation of $\mu_0$ restricted
to $D_n$. The decomposition requires $\mathcal{O}(n^3)$ operations unless 
the matrix $C_n$ has specific types of sparsity, hence obtaining this 
white noise representation may be infeasible when $n$ is
very large. See \cite{williams2006gaussian} for an overview. Notice that this construction only refers to a finite-dimensional or a discretized Gaussian process and does not have an infinite-dimensional analogue. 

\subsubsection{Factorisations of precision operator.}
\label{ssec:prec}
Consider the
Gaussian measure $\mu_0=\sN(0,\cC)$ on $X=L^2(D;\R),$ the space of square integrable functions on $D$, where the inverse covariance operator, $\cL$, is some
    densely defined differential operator $\cL$ on $X$. The locality
of the differential operator is a form of infinite dimensional sparsity,  
and certain finite dimensional approximations of the operator, e.g.  finite elements or finite difference methods, result in sparse matrices. In fact, there is a close link between Gaussian Markov random fields and Gaussian processes with differential precision operators, see  \cite{papaspiliopoulos2012nonparametric}; therefore, there are canonical ways to discretize $u$ and its distribution so that the resultant finite-dimensional process is a Markov random field.     
Suppose further that $\cL$ may be factorised as $\cL = \mathcal{A}^*\mathcal{A}$, where $\mathcal{A}$ is itself a differential operator. Then samples
    $u\sim\mu_0$ can be generated by solving the stochastic PDE (SPDE) 
\begin{align}
\label{eq:spde_general}
\mathcal{A} u = \xi,
\end{align}
which provides the white noise representation by taking  $T$ to be the solution mapping
$\xi\mapsto u$. $\Xi$ will be larger than $X$
(see the Appendix)  but
its image under $T$ will be contained in $X$
as $T$ is a smoothing operator. \cite{matern_spde} systematise the above construction when the Gaussian process is specified via taking its covariance function to be in the Mat\'ern family, see \Cref{ex:whittlematern} below. When a sparsity-preserving discretisation is applied to obtain a matrix $A_n$ approximating $\mathcal{A}$, so that $A_n^*A_n$ approximates $\mathcal{L}$, the
white noise representation of the finite-dimensional approximation is given by $(\R^n,\sN(0,I),T(\xi)=A_n^{-1} \xi)$.  Fast PDE
solvers can be utilised to evaluate $T$ efficiently. In contrast to \Cref{ssec:chol} this is based on a factorisation of
the precision as opposed to the covariance matrix. Simulation and computations for finite-dimensional Gaussian Markov random fields based on Cholesky factorisation of the precision matrix was systematised in \cite{rue:book}.

\subsubsection{Karhunen-Lo\`eve expansion.}
\label{ssec:KL}
The properties of $\cC$ mean that it admits a complete orthonormal basis
of eigenvectors $\{\varphi_j\}_{j\geq 1}$ for $X$ with corresponding
non-negative and summable eigenvalues $\{\lambda_j\}_{j\geq 1}$; this
gives rise to natural spectral methods where functions in $X$ are
represented via expansions in this basis, and approximations
may be made by truncating such expansions. As a
consequence of the Karhunen-Lo\`eve theorem
$\mu_0=\sN(0,\cC)$ is equal to the law of the random variable $u$ defined by
\[
u = \sum_{j=1}^\infty \sqrt{\lambda_j}\xi_j\varphi_j,\quad \xi_j\iid \sN(0,1).
\]
Thus, the white noise representation of $\mu_0$ is obtained by taking 
\[
T(\xi) := \sum_{j=1}^\infty \sqrt{\lambda_j}\xi_j\varphi_j.
\]
 This series-based construction and white noise representation will be
 the basis of the methodology for non-Gaussian priors in \Cref{sec:mcmc}.  This approach can be related to that of \Cref{ssec:prec}, since the $\varphi_j$ are eigenfunctions of $\cL$
and $\lambda_j^{-1}$ the corresponding eigenvalues, hence  this method can be seen as a special case of a factorisation of the precision operator,  wherein a spectral method is used for the evaluation of $T$.

\subsection{Example: Whittle-Mat\'ern process}
\label{ex:whittlematern}
This process serves to illustrate all of the above constructions, and will be the main building block in  hierarchical priors. We start by introducing the process through its
covariance function
\[
c(x,x') 
= \sigma^2 \frac{2^{1-\beta}}{\Gamma(\beta)}\left(\tau|x-x'|\right)^\beta K_\beta\left(\tau |x-x'|\right),\quad x,x' \in \R^d,
\]
where $\sigma,\tau,\beta>0$ are scalar parameters representing
standard deviation, inverse length-scale and regularity respectively.
Here $\Gamma$ is the gamma function and $K_\beta$ is the modified
Bessel function of the second kind of order $\beta$. If $\mu_0 =
\GP(0,c(x,x'))$ 
and $u \sim \mu_0$, then almost-surely $u$ has $s$ Sobolev and
H\"older derivatives for any $s < \beta$. 
This can be used to define a  Gaussian process over the whole of $\R^d$, but the covariance
can be restricted to a bounded open subset $D.$

The Cholesky factorisation of the covariance matrix can be used to
obtain a white noise representation 
if we restrict to a finite subset of points $\{x_j\}_{j=1}^n$.
We can alternatively obtain a white noise
representation of the process everywhere on $D$ by using a factorisation of the
precision operator which, for this covariance on the whole of $\R^d$,
has the form
\begin{align}
\label{eq:matern_spde2}
\mathcal{L} & = \sigma^{-2} \tau^{-2\beta}q(\beta)^{-1}(\tau^2
I-\Delta)^{(\beta+d/2)},\quad q(\beta) = \frac{2^d \pi^{d/2}\Gamma(\beta+d/2)}{\Gamma(\beta)},
\end{align}
where $\Delta$ is the second-order differential Laplace operator. 
In \cite{matern_spde} it is noted that the square root of $\mathcal{L}$ is
\begin{align}
\label{eq:matern_spde}
\mathcal{A} & = \sigma^{-1} \tau^{-\beta}q(\beta)^{-1/2}(\tau^2
I-\Delta)^{(\beta+d/2)/2}.
\end{align}
We thus have a factorisation of the precision, and this may be used
to define a white noise representation of the Gaussian. The method can
be implemented on a finite domain $D \subset \R^d$ by means of finite difference or finite
element methods  and 
choosing appropriate boundary conditions when $(\beta+d/2)/2$ is an integer. 
Note, however, that the the boundary conditions may 
modify the covariance structure near the boundary.\footnote{A simple method to ameliorate these effects is to perform sampling on a larger domain $D^*\supset D$ so that samples restricted to $D$ are approximately stationary. More complex methods have also been considered, for example by optimal choice of a constant Robin boundary condition, as in  \cite{roininen2014whittle}, or of a variable Robin boundary condition combined with variance normalisation, 
as in \cite{daon2016mitigating}.}
When the exponent of the differential operator in $\mathcal{A}$ is not an integer, its action may be defined through the Fourier transform $\mathcal{F}$, i.e.
\[
\mathcal{F}(\mathcal{A}u)(\omega) = \sigma^{-1} \tau^{-\beta}q(\beta)^{-1/2}(\tau^2
+ |\omega|^2)^{(\beta+d/2)/2}.
\]
This leads to spectral methods and the Karhunen-Lo\`eve white noise
representation.  Suppose $D \subset\R^d$ is a bounded rectangle, 
and homogeneous Neumann or Dirichlet boundary condition are applied. Then the eigenvectors of $\mathcal{A}$ are known analytically, and given by Fourier basis functions. For example, if $D = (0,1)$ and homogeneous Neumann boundary conditions are assumed, we have
\[
\mathcal{C}\varphi_j = \lambda_j\varphi_j,\quad \varphi_j(x) = \sqrt{2}\cos(j\pi  x),\quad \lambda_j = \sigma^2\tau^\beta q(\beta)(\tau^2 + \pi^2 j^2)^{-\beta-d/2}.
\]
The Karhunen-Lo\`eve expansion may then be efficiently implemented numerically using the fast Fourier transform.

\section{Series expansions and level-set priors}
\label{sec:mcmc}

We consider priors $\mu_0(\dee u)$ based on series expansions with non-Gaussian coefficients, such as uniform, Besov and stable processes, and level-set transformations of a Gaussian process. For these processes we show how to obtain the  white noise representation $(\Xi,\nu_0,T)$ of $\mu_0$. Recall that $\nu^y(\dee \xi|\theta)$ denotes the posterior distribution of the transformed latent state; we will drop $\theta$ from the formulae and re-introduce it in \Cref{sec:hierarchical} when we treat hierarchical priors. 
We develop dimension-robust MCMC samplers for $\nu^y$, and then illustrate their
efficiency by means of several numerical experiments.

\subsection{Robust algorithms}
\label{ssec:rob3}

Here we show explicitly how to adapt the pCN and $\infty$-MALA algorithms so that
they apply to the transformed posterior $\nu^y$. In doing so we're defining
dimension robust algorithms for inverse problems with non-Gaussian
priors. Other dimension-robust algorithms for Gaussian priors can be adapted to 
non-Gaussian priors in a similar fashion: in this Section we also show results for 
an adaptation of $\infty$-HMC to the non-Gaussian setting. 
Adapting pCN is trivial and yields what we call whitened pCN (wpCN). 

\begin{algorithm}
\caption{Whitened Preconditioned Crank-Nicolson (wpCN)}
\label{alg:wpcn}
\begin{algorithmic}[1]
\State Fix $\beta \in (0,1]$. Choose initial state $\xi^{(0)}\in \Xi$.
\For{$k=0,\ldots,K-1$}
\State  Propose $\hat{\xi}^{(k)} = (1-\beta^2)^{\frac{1}{2}}\xi^{(k)} + \beta\zeta^{(k)},\quad\zeta^{(k)} \sim \sN(0,I)$.
\State \parbox[t]{\dimexpr\linewidth-\algorithmicindent}{Set $\xi^{(k+1)} = \hat{\xi}^{(k)}$ with probability
\[
\min\Big\{1,\exp\big(\Phi(T(\xi^{(k)});y) - \Phi(T(\hat{\xi}^{(k)});y)\big)\Big\}
\]
or else set $\xi^{(k+1)} = \xi^{(k)}$.\strut}
\EndFor
\State\Return $\{T(\xi^{(k)})\}_{k=0}^{K}$.
\end{algorithmic}
\end{algorithm}

Adaptation of $\infty$-MALA is a little trickier since it requires computing gradients. Implementing the algorithm when $\Xi$ is finite-dimensional is straightforward, all derivatives can be computed in the familiar way. The norm and the inner product that show up in \Cref{alg:wmala} are the familiar Euclidean norm and dot product in this finite-dimensional setting. Note that  $T$ is differentiable for the 
uniform, Besov and stable priors described below, and the gradients are available 
explicitly, but not for the level-set prior.
The whitened $\infty$-MALA (w$\infty$-MALA) we present below as \Cref{alg:wmala} is conceptually valid even when $\Xi$ is infinite-dimensional. It is in this infinite-dimensional vesion that a little care is needed to define the quantities involved properly, and the Appendix gives details on how to construct and interpret the gradients, norms and inner products appearing in \Cref{alg:wmala}.  

\begin{algorithm}
\caption{Whitened $\infty$-MALA (w$\infty$-MALA)}
\label{alg:wmala}
\begin{algorithmic}[1]
\State Fix $h\in (0,4]$ and define $\beta = 4\sqrt{h}/(4+h) \in (0,1]$. Define $\Psi(\xi) = \Phi(T(\xi))$, with gradient evaluated at $\xi$,   $D \Psi(\xi)$.  Choose initial state $\xi^{(0)}\in \Xi$.
\For{$k=0,\ldots,K-1$}
\State  Propose $\hat{\xi}^{(k)} = (1-\beta^2)^{\frac{1}{2}}\xi^{(k)} + \beta\left(\zeta^{(k)} - \frac{\sqrt{h}}{2}D \Psi(\xi^{(k)}) \right),\quad\zeta^{(k)} \sim \sN(0,I)$.
\State \parbox[t]{\dimexpr\linewidth-\algorithmicindent}{Set $\xi^{(k+1)} = \hat{\xi}^{(k)}$ with probability
\[
\min\Big\{1,\exp\big(I(\xi^{(k)},\hat\xi^{(k)}) - I(\hat\xi^{(k)},\xi^{(k)})\big)\Big\}
\]
where we have defined
\[
I(\xi,\xi') := \Psi(\xi) + \frac{h}{8}||D \Psi (\xi) )||^2 + \frac{\sqrt{h}}{2}\left\la  D\Psi (\xi),\frac{\xi'-\sqrt{1-\beta^2}\xi}{\beta}\right\ra
\]
or else set $\xi^{(k+1)} = \xi^{(k)}$.\strut}
\EndFor
\State\Return $\{T(\xi^{(k)})\}_{k=0}^{K}$.
\end{algorithmic}
\end{algorithm}

\subsection{White noise representation of series expansions}
\label{ssec:white3}

The common structure for the priors we consider in this Section is their representation as series expansions with non-Gaussian coefficients. The construction
of these priors is inspired by the Karhunen-Lo\`eve expansion for
Gaussian processes, as in \Cref{ssec:KL}, except the basis does not necessarily correspond
to the eigenbasis of a given covariance operator, and the randomness introduced
to each mode is not necessarily Gaussian. Another approach to define
families of non-Gaussian distributions on function space is through
the SPDE \cref{eq:spde_general} when the white noise is non-Gaussian;
we do not consider this approach explicitly, though the ideas
discussed can be applied in such cases too.
In what follows, the prior $\mu_0$ will be given by the law of the $X$-valued random variable defined by
\begin{align}
\label{eq:series_general}
u = m + \sum_{j=1}^\infty \rho_j\zeta_j\varphi_j,
\end{align}
where $\rho = \{\rho_j\}_{j\geq 1}$ is a deterministic real-valued sequence, $\zeta_j \sim G_j(\dee \zeta)$  are independent,  $\{\varphi_j\}_{j\geq 1}$ is a deterministic $X$-valued sequence, and $m \in X$.  The white noise representation for any such prior is
\[
T(\xi) = m + \sum_{j=1}^\infty \rho_j\Lambda_j(\xi_j)\varphi_j,
\]
where $\Lambda_j(\xi)$ are transformations of standard Gaussians so that $\Lambda_j(\xi) \sim G_j$. 
We now describe a number of families of non-Gaussian distributions and obtain a transformations $\Lambda_j(\cdot)$. Throughout the Section $F$ will denote the cumulative distribution function of a standard normal random variable.

\subsubsection{Uniform priors.}
Here we work with the uniform prior; see \cite{lecturenotes} for a historical
review.  
Let $D\subseteq \R^d$ be a bounded open subset, let $\{\varphi_j\}_{j\geq 1}\subseteq L^\infty(D)$, and define $X$ to be the closure of the linear span of this sequence in $L^\infty(D)$. Assume that $\rho \in \ell^1$, $m \in X$, and let $\zeta_j \iid \sU(-1,1)$. Then $\Lambda(z) = 2F(z) - 1$ and  let $\xi \sim \sN(0,1)$;   it is
elementary that 
$\Lambda(\xi) \sim \sU(-1,1)$.

\subsubsection{Besov priors.}
\label{ssec:besov_prior}
Besov priors were introduced in \cite{Lassas200987} and analysed further in \cite{DHS12}. They generalise Gaussian priors by allowing for control over the weight of their tails. Let $D\subseteq\R^d$ be a bounded open domain, and define $X=L^2(D)$. Let $\{\varphi_j\}_{j\geq 1}$ be a basis for $X$, and choose $m \in X$. Given $s > 0$ and $q\geq 1$, define the Banach space $X^{s,q}\subseteq X$ through the norm
\[
\|u\|_{X^{s,q}}^q = \sum_{j=1}^\infty j^{sq/d+q/2-1}|u_j|^q\quad\text{where}\quad u = \sum_{j=1}^\infty u_j\varphi_j.
\]
If $D = \mathbb{T}^d$ is the torus, and $\{\varphi_j\}_{j=1}^\infty$ is chosen to be an $r$-regular wavelet basis for $X$ with $r>s$, then $X^{s,q}$ is the Besov space $B^s_{qq}$ \cite{Lassas200987}.

The series based construction \cref{eq:series_general} can be used to produce probability measures with strong links to the above spaces. Given $s,q$ as above and $\kappa > 0$, define the sequence $\{\rho_j\}_{j\geq 1}$ by
\[
\rho_j = \kappa^{-\frac{1}{q}}j^{-(\frac{s}{d}+\frac{1}{2}-\frac{1}{q})}.
\]
Let $\{\zeta_j\}_{j\geq 1}$ be an i.i.d. sequence of draws from the distribution defined via the density
\[
\pi_q(x) \propto \exp\left(-\frac{1}{2}|x|^q\right),
\]
and let $m \in X$. The measure $\mu_0$ is a $(\kappa,B^s_{qq})$
measure in the sense of \cite{DHS12} and, when $m=0$, it formally
has Lebesgue density proportional to
$\exp\left(-\frac{\kappa}{2}\|u\|_{X^{s,q}}^q\right)$. The cases $q=1$
are of particular interest since they allow for discretisation
invariant edge-preserving Bayesian inversion; this is in contrast to
total variation priors which are often used in classical inversion for
edge-preservation as in \cite{Lassas200987}. MAP estimation using these
priors is well-defined and corresponds to Besov regularised
optimisation as in \cite{agapiou2017sparsity}. These methods may be viewed
as Bayesian and infinite-dimensional analogues of the lasso.  

We construct a white noise representation of $\mu_0$. In order to to this, we first write down a method for sampling the scalar distribution $\pi_q$. We use the method of \cite{nardon2009simulation}: a sample $\zeta \sim \pi_q$ can be produced as
\[
\zeta = 2^{1/q}B\cdot G^{1/q},\quad B \sim \mathsf{Bernoulli}(1/2),\quad G\sim \mathsf{Gamma}(1/q,1)
\]
where $B$ and $G$ are independent. The proposed white noise transformation is based on the transformation 
\[
\Lambda(\xi) = 2^{1/q}\sgn(\xi)\Big(\gamma_{1/q}^{-1}\big(2F(|\xi|)-1\big)\Big)^{1/q}
\]
where $\gamma_{1/q}$ is the normalised lower incomplete gamma function:
\[
\gamma_{1/q}(z) = \frac{1}{\Gamma(1/q)}\int_0^z t^{1/q-1}e^{-t}\,\dee t.
\]
It can be checked that this transformation of a standard Gaussian has the desired distribution. 

\begin{remark}
A similar white noise representation to the above is provided in
\cite{marzouk}, in the cases where $q=1$ and the sum
\cref{eq:series_general} is truncated after a finite number of
terms. A randomise-the-optimise approach is used to
sample the posterior rather than MCMC, which should extend easily to
the cases $q>1$ considered here. 
\end{remark}

\subsubsection{Stable priors.}
Stable priors  are introduced and analysed in
\cite{sullivan2016well} in the context of Bayesian inversion. Their
intersection with Besov distributions is precisely the set of Gaussian
distributions, and they arise by assuming that each of the random
variables $\zeta_j$ in \cref{eq:series_general} is a stable random variable. The set of stable distributions on $\R$ can be interpreted as the set of distributions that are limits in central limit theorems, and may be defined via their characteristic function. Let $\alpha \in (0,2]$, $\beta \in [-1,1]$, $\gamma \in (0,\infty)$ and $\delta \in \R$ be scalar parameters, representing stability, skewness, scale and location respectively. We will say that a real-valued random variable $\zeta$ has stable distribution $\mathsf{S}(\alpha,\beta,\gamma,\delta)$ if, for each $t \in \R$,
\[
\mathbb{E}\big(e^{-it\zeta}\big) =\begin{cases}
\exp\big(it\delta - |\gamma t|^\alpha[1+i\beta\tan\big(\frac{\pi\alpha}{2}\big)\mathrm{sgn}(t)(|\gamma t|^{1-\alpha}-1)]\big) & \alpha \neq 1\\
\exp\big(it\delta - |\gamma t|[1+i\beta\frac{2}{\pi}\mathrm{sgn}(t)\log(\gamma|t|)]\big) & \alpha = 1.
\end{cases}
\]
The Lebesgue density of a $\mathsf{S}(\alpha,\beta,\gamma,\delta)$
distribution is typically not expressible analytically, however it is
known these distributions are unimodal and possess $\lceil \alpha -1
\rceil$ moments; $\alpha = 2$ corresponds to the Gaussian
distributions, and  $\alpha = 1$ to the Cauchy.

Let $D\subseteq\R^d$ be a bounded open domain, $X = L^2(D)$, and let $\{\varphi_j\}_{j\geq 1}$ be a normalised basis for $X$. Given $\alpha \in (0,2]$ and sequences $\beta = \{\beta_j\}\subseteq [-1,1]^\infty$, $\gamma = \{\gamma_j\} \subseteq (0,\infty)^\infty$ and $\delta = \{\delta_j\}\subseteq \R^\infty$, and 
\[
\zeta_j \sim \mathsf{S}(\alpha,\beta_j,\gamma_j,\delta_j)\quad\text{independent.}
\]
Define also $\rho_j=1$ for each $j$, and let $m\in X$. 

For the white noise representation we will  in general require two independent $\sN(0,1)$ random variables to construct a single $\zeta_j$. The transformation is based on the method presented by \cite{chambers1976method}, which is a generalisation of the Box-Muller transform for sampling Gaussian random variables. For $\xi,\xi' \iid \sN(0,1)$ we define  
\begin{align*}
\small
\Lambda(\xi,\xi';\alpha,\beta,\gamma,\delta) = 
\begin{cases}
\delta + \gamma(1+\tau^2)^{\frac{1}{2\alpha}}\frac{\sin(\alpha(U(\xi)+\theta))}{\cos(U(\xi))^{1/\alpha}}\left\{\frac{\cos(U(\xi) - \alpha(U(\xi)+\theta))}{W(\xi')}\right\}^{\frac{1-\alpha}{\alpha}} &\alpha \neq 1\\
\delta + \tau + \frac{\gamma}{\theta}\left\{\left(\frac{\pi}{2} + \beta U(\xi)\right)\tan(U(\xi)) - \beta\log\left(\frac{\pi W(\xi')\cos(U(\xi))}{\pi + 2\beta U(\xi)}\right)\right\} & \alpha = 1
\end{cases}
\end{align*}
where
\[
\tau = 
\begin{cases}
-\beta\tan\left(\frac{\pi\alpha}{2}\right) & \alpha \neq 1\\
\frac{2}{\pi}\beta\gamma\log(\gamma) & \alpha=1,
\end{cases}
\quad \theta =
\begin{cases}
\frac{1}{\alpha}\tan^{-1}(-\tau) & \alpha \neq 1\\
\frac{\pi}{2} & \alpha=1.
\end{cases}
\]
The white noise representation takes $\nu_0 = \sN(0,I)\times \sN(0,I)$, and the transformation is 
\[
T(\xi,\xi') = \sum_{j=1}^\infty \Lambda(\xi_j,\xi_j';\alpha,\beta_j,\gamma_j,\delta_j)\varphi_j.
\]

\subsection{White noise representation of level-set priors}
\label{sssec:lvl_prior}

A large class of inverse problems involve the recovery of a piecewise
constant function. This class includes classification problems as in
\Cref{ex:inv_class} 
where the unknown function is a mapping from the set of data points to a discrete set of classes. It also includes PDE-based inversion as in \Cref{ex:inv_pde}, which we analyze in detail in \Cref{ssec:darcy}: here the unknown permeability may be approximately piecewise constant, with the different values corresponding to the permeability of different materials. The key part of such inversion is the recovery of the interfaces separating the different classes. Level-set methods are a popular choice of methods for inverse interface problems, as they require no prior knowledge or assumption on the topology of the different classes, see \cite{Osher1988, Santosa1996}. 
The Bayesian level-set  and hierarchical Bayesian level-set methods were recently introduced in  \cite{levelset} and \cite{levelsethier} respectively to allow for uncertainty quantification in inverse interface problems.

The idea of level-set methods is to create a piecewise
constant field by thresholding a continuous field. Let
$D\subseteq\R^d$ be a bounded open domain, and define $X =
L^\infty(D;\R)$, the space of bounded measurable $\R$-valued functions on $D$. Choose classes $\kappa_1,\ldots,\kappa_k \in \R$ and thresholding levels $c_1<\ldots<c_{k-1} \in \R$, and define 
\begin{align}
\label{eq:level_ordered}
u(x) = 
\begin{cases}
\kappa_1 & v(x) \leq c_1\\
\kappa_2 & c_1 < v(x) \leq c_2\\
\vdots &\\
\kappa_{k-1} & c_{k-2} < v(x) \leq c_{k-1}\\
\kappa_k & c_{k-1} < v(x),
\end{cases}
\end{align}
for $v$ a  continuous function. Hence, given a measure $\lambda_0(\dee v)$, e.g. such a Gaussian measure, implies a measure $\mu_0(\dee u)$, that concentrates on
piecewise constant functions. To obtain a white noise representation
of $\mu_0$ we first obtain the one for $\lambda_0$, say $(\Xi,\nu_0,T_0)$, using a
method as described in \Cref{ssec:white2,ssec:white3}, 
and then $(\Xi,\nu_0,T)$, with $T$ the composition of $T_0$ with the level-set transformation defined above,  provides a white noise representation of
$\mu_0$. In our examples in this paper $\lambda_0$ is always Gaussian. 

In  the above method is there is an ordering of
the classes: arbitrary classes cannot share an interface, and so, for
example, a triple-junction cannot be formed.  Vector level-set
methods allow one to get around this restriction, in exchange for
increasing the dimension of the unknown field. One such method is that of \cite{hu2015multi}, in which we $k$ continuous functions $v_1,\ldots,v_k$, and define
\begin{align}
\label{eq:multiclass_threshold}
u(x)  = \mathds{1}_{r(x;v)},\quad r(x;v) = \underset{r=1,\ldots,k}{\mathrm{arg}\max}\;v_r(x),
\end{align}
where $\{\mathds{1}_r\}_{r=1}^k$ denotes the standard basis for $\R^k$.

In familiar terms from Statistics, the model with one latent continuous function is appropriate for ordinal data whereas the one with $k$ is for categorical data. In fact, the models discussed above are closely related to probit models. To see the connection let $S(v)$ denote the map from $v$ to the discrete values, and consider $k=2$ levels for simplicity with  $\kappa_1 = 1,\kappa_2 = -1$. A simplistic (but often used for computational convience) way to model categorical data is via the regression, as in \Cref{ex:reg} 
\[
y_j = S\bigl(v(x_j)\bigr)+\eta_j,\quad\eta_j \sim \sN(0,\gamma).
\]
A more satisfctory way to model such data is to  model the noise as additive with respect to $v$ and not $u$, so that 
\[
y_j = S\bigl(v(x_j)+\eta_j\bigr),\quad\eta_j \sim \sN(0,\gamma),
\]
then we obtain the negative log-likelihood
\[
\Phi(v;y) = -\sum_{j=1}^J \log F\big(v(x_j)y_j/\gamma\big),
\]
where $F$ is the standard normal CDF, which is precisely  the probit likelihood function \cite{williams2006gaussian}.

\subsection{Simulation experiments}
\label{sub:sim-ng}

We first compare the wpCN algorithm and w$\infty$-MALA
with two random walk Metropolis (RWM) algorithms on a problem with a Besov prior. 
The prior is of the form described in \Cref{ssec:besov_prior}, with $q=1$, $s=1$, $\kappa = 0.1$ and $d=2$, and a Fourier basis is used for the expansion; specifically we take $m = 0$, $\rho_i = (k_1^2+k_2^2)^{-1}$ and $\varphi_i(x,y) = 2\cos(k_1\pi x)\cos(k_2\pi y)$, where we have enumerated $k_1^2+k_2^2\asymp i$. The model is basic
linear regression as described in \Cref{ex:reg};  the $\{x_j\}$ comprise
$16$ points on a uniform grid in the domain $(0,1)^2$, and the output data $\{y_j\}$
are found by perturbing $\{u(x_j)\}$ with i.i.d. white noise with standard deviation 
$0.1.$ The true field is drawn from the prior distribution. For practical implementation we truncate the series at size $N$; in the experiments we have considered various sizes. The MCMC sampling is done on the series coefficients but at each iteration of the algorithm we also compute the approximate (due to truncation of series) point evaluations $u(x_j), j \in Z$ to compute the likelihood.  
The two RWM algorithms differ in the distribution of their proposal jumps. Each RWM proposal is of the form $u\mapsto u + \beta \zeta$ for some $\beta > 0$, and $\zeta$ is a centred random variable. We consider both the case where $\zeta \sim \sN(0,I)$ is Gaussian white noise, and the case where $\zeta$ is drawn from the prior distribution.
Results from this simulation experiment were shown in \Cref{fig:pcn_vs_rwm}, which clearly shows that for wpCN and w$\infty$-MALA algorithms these curves are stable under refinement of the discretisation; in contrast for the RWM algorithms the curves shift with changing discretisation and, in particular, the proposal size allowable for given acceptance probability decreases with dimension.

\begin{figure}
\begin{center}
\includegraphics[scale=0.4]{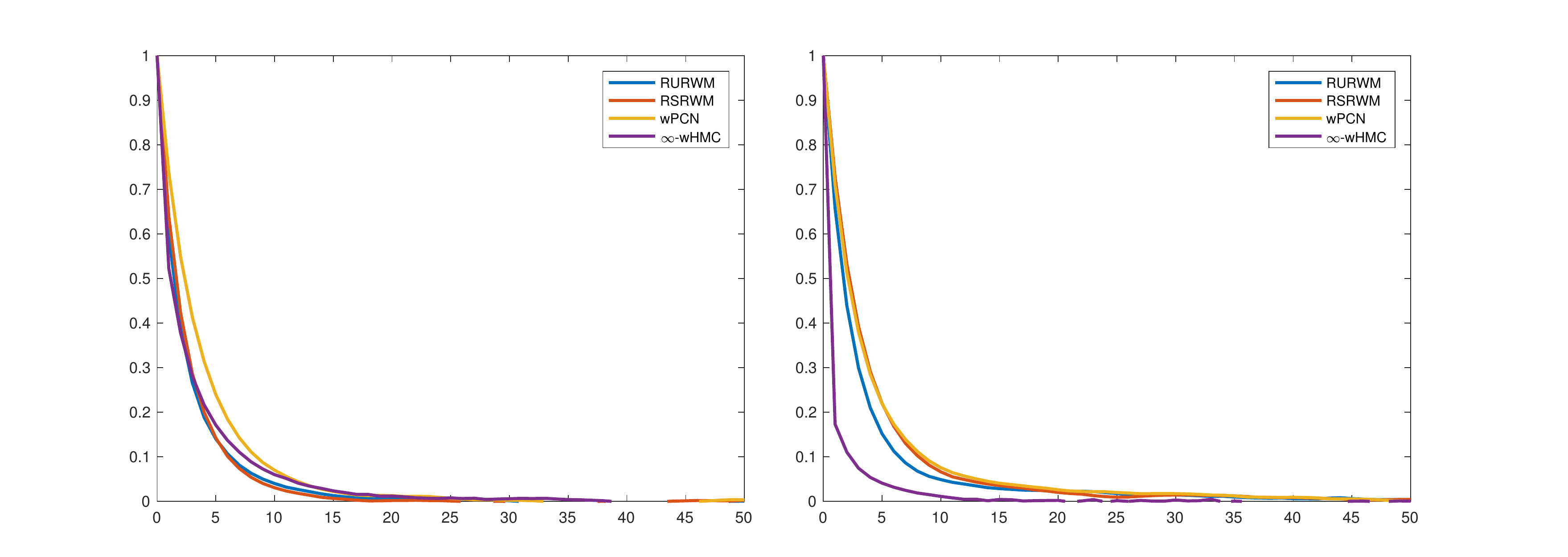}
\end{center}
\caption{Autocorrelations for $||u||$, the $L^2$-norm of the latent function, for different MCMC samplers using the convolution 
forward model, with 8 observations (left) and 32 observations (right).}
\label{fig:compare_bhe}
\end{figure}

  We now compare the generic methodology of this paper to the methodology presented in 
\cite{vollmer2015dimension} for uniform priors.
We compare with two methods that are designed to be dimension-robust for this specific model, the reflected uniform and Gaussian random walk proposals, referred to as 
RURWM and RSRWM respectively and derived in \cite{vollmer2015dimension}. We emphasize that the methods we promote here are agnostic to the details of the models and only require the white noise representation of the prior. 
We study a version of \Cref{ex:reg} where the noise observations are not of $u$ but $K(u)$, where $K(u)(x) = \sum_i e^{-0.1 i } \rho_i \zeta_i \phi_j(x)$, for $x \in (0,1)$ and $u(x) = \sum_i \rho_i \zeta_i \phi_i(x)$; the observations are $y_j = K(u)(x_j) + \eta_j$, with $x_j$'s equally spaced on the domain $(0,1)$. A uniform prior of the form \cref{eq:series_general} is used, with $m = 2$, $\rho_i = 1/i^2$ and $\varphi_i(x) = \sqrt{2}\cos(i\pi x)$ for each $i \geq 1$. Jump size parameters are chosen such that the acceptance rate is roughly $30\%$ for wpCN, RURWM and RSRWM, and roughly $60\%$ for w$\infty$-HMC. The true field is generated on a mesh of $2^{12}$ points, and observations are corrupted with Gaussian noise with standard deviation such that the average relative error is approximately $4\%$. All methods considered sample  $2^{10}$ coefficients - that is, they all work on the frequency domain and sample a very high-frequency approximation of $u$. 
Autocorrelations of the sampled $||u||$, the $L^2$ norm of the latent function, for different numbers of observation points are shown in \Cref{fig:compare_bhe}. With the weaker likelihood derived from 8 observations, the results demonstrate that the wpCN and 
w$\infty$-HMC algorithms behave similarly to the existing ad hoc methods. With a stronger likelihood derived from 32 observations, we see the advantage of using the likelihood-informed proposal of w$\infty$-HMC.

\section{Hierarchical priors}
\label{sec:hierarchical}

Taking into account parameter uncertainty, we now define $\mu_0$ as a
hierarchical prior disintegrated as $\mu_0(\dee u,
\dee \theta) = \mu_0(\dee u | \theta) \pi_0(\theta) \dee \theta$, with 
$\theta$ a vector of parameters that define it and 
where we assume that the marginal prior distribution on the parameters
$\theta$ admits Lebesgue density $\pi_0$. Note the conditional distribution $\mu_0(\dee u|\theta)$ may concentrate on different subsets of $X$ for different $\theta$, such as sets of functions with a specific regularity. The
target posterior is now 
\[
\mu^y (\dee u, \dee \theta) \propto \exp(-\Phi(u;y))  \mu_0(\dee u | \theta) \pi_0(\theta) \dee \theta\,
\]
and the aim is to design dimension-robust algorithms for sampling such distributions.

\subsection{Robust algorithms}

Joint posteriors of latent states and parameters are typically sampled using a Metropolis-within-Gibbs algorithm
that samples iteratively from the two conditional distributions: 
\begin{enumerate}
\item $\mu^y (\dee u |
  \theta) \propto \exp(-\Phi(u;y))  \mu_0(\dee u | \theta)$ 
\item $\mu^y (\theta | u) \propto
  \mu_0(\dee u | \theta) \pi_0(\theta)$. 
\end{enumerate}
The simulation problem in step 1 is analogous to those we have discussed in previous sections; that in step 2 is typically a low-dimensional sampling problem that can be performed in more or less straighforward way. The above formulation  is an instance of what in Bayesian hierarchical models is
commonly called a centred algorithm or equivalently a
Metropolis-within-Gibbs algorithm based on a centred parameterisation
of the model, see \cite{papaspiliopoulos2007general} for details and
overview. 

It is well-documented that such component-wise updating schemes will
mix poorly whenever the measures $\mu_0(\dee u | \theta)$, for
different $\theta$'s, are very different, say with large total
variation distance for small perturbations of $\theta$. In fact, in
many applications where $X$ is infinite-dimensional these measures are
mutually singular and the centred algorithm is reducible, i.e., it
would never move from its initial values; for the first treatment of this problem see \cite{roberts2001inference}, for an overview see \cite{papaspiliopoulos2007general} and a theoretical analysis for linear hierarchical Gaussian inverse problems see \cite{agapiou2014analysis}. 
A generic solution to this pathology is to work with a non-centred
parameterisation of the hierarchical model, which is defined to be one
under which a transformed latent state and the parameters are a priori
independent. The Metropolis-within-Gibbs algorithm that targets the
corresponding posterior is termed non-centred algorithm.  The
motivation behind this parameterisation, especially 
for infinite-dimensional models,  is the following: if the data are
not infinitely informative about $u$, hence a likelihood function
exists, sets that have probability 1 under the prior also do
under the posterior; hence if under the prior latent states and
parameters are not perfectly dependent they will also not be under the
posterior and the non-centred algorithm will be ergodic.

The developments of the previous two sections readily provide
non-centred parameterisations of the hierarchical Bayesian inverse
problem since the Gaussian white noise latent state $\xi \sim \nu_0$, and parameters 
$\theta$ defining the map $u = T(\xi,\theta)$ are
a priori independent. The likelihood thus depends on both $\xi$ and $\theta$ and the
posterior takes the form 
\[
\nu^y(\dee \xi, \dee \theta) \propto \exp(-\Phi(T(\xi,\theta);y))
\nu_0(\dee \xi) \pi_0(\theta) \dee \theta.
\]
By working in variables $(\xi,\theta)$ rather than $(u,\theta)$ we completely
avoid lack of robustness arising from mutual singularity, by  applying Metropolis-within-Gibbs to 
the variables $(\xi,\theta)$ rather than $(u,\theta)$. Furthermore, by sampling
$\nu^y(d\xi|\theta)$ using a dimension-robust sampler as in the previous
section we construct an overall methodology which is dimension-robust.
The method is provided as pseudocode in \Cref{alg:mwg}.
\begin{algorithm}
\caption{Non-centred Preconditioned Crank-Nicolson Within Gibbs}
\label{alg:mwg}
\begin{algorithmic}[1]
\State Fix $\beta \in (0,1]$. Choose initial state $(\xi^{(0)},\theta^{(0)}) \in \Xi$.
\For{$k=0,\ldots,K-1$}
\State  Propose $\hat{\xi}^{(k)} = (1-\beta^2)^{\frac{1}{2}}\xi^{(k)} + \beta\zeta^{(k)},\quad\zeta^{(k)} \sim \sN(0,\cC)$.
\State \parbox[t]{\dimexpr\linewidth-\algorithmicindent}{Set $\xi^{(k+1)} = \hat{\xi}_1^{(k)}$ with probability
\[
\min\left\{1,\exp\left(\Phi(T(\xi^{(k)},\theta^{(k)});y) - \Phi(T(\hat{\xi}^{(k)},\theta^{(k)});y)\right)\right\}
\]
or else set $\xi^{(k+1)} = \xi^{(k)}$.\strut}
\State Propose $\hat{\theta}^{(k)} \sim q(\theta^{(k)},\cdot)$.
\State \parbox[t]{\dimexpr\linewidth-\algorithmicindent}{Set $\theta^{(k+1)} = \hat{\theta}^{(k)}$ with probability
\[
\min\left\{1,\exp\left(\Phi(T(\xi^{(k+1)},\theta^{(k)});y) - \Phi(T(\xi^{(k+1)},\hat{\theta}^{(k)});y)\right)\frac{q(\hat{\theta}^{(k)},\theta^{(k)})}{q(\theta^{(k)},\hat{\theta}^{(k)})}\frac{\pi_0(\hat{\theta}^{(k)})}{\pi_0(\theta^{(k)})}\right\}
\]
or else set $\theta^{(k+1)} = \theta^{(k)}$.\strut}
\EndFor
\State\Return $\{T(\xi^{(k)},\theta^{(k)})\}_{k=0}^{K}$.
\end{algorithmic}
\end{algorithm}

\subsection{Simulation experiments}

\label{ssec:darcy}
We consider hierarchical level-set priors for two PDE inversion problems; one is recovering the permeability field using pressure measurements and is described in \Cref{ex:inv_pde}, the second is a problem from medical imaging, specifically Electrical Impedance Tomography, and it is described in detail in this section.

The permeability field we aim to recover is shown in \Cref{fig:gwf_truth}, together with the spatial locations of the 36 points at which pressure is measured; pressure is a highly non-linear map of the latent permeability field. Observations are corrupted by Gaussian white noise with standard deviation 0.05, resulting in an average relative error of $7.5\%$.
We first consider a level set prior in which we threshold a Whittle-Mat\'ern field at two levels. We then consider two hierarchical priors in which the inverse length-scale parameter $\tau$ of the underlying Gaussian field is treated as a parameter, first using an ad-hoc sampling method introduced in \cite{levelsethier}, and then using a non-centred method as considered in this article. The method of \cite{levelsethier} can be considered both centred, in that it maintains correlations between the field and hyperparameter under the prior, and non-centred, in that the likelihood is modified to explicitly depend on the hyperparameter in such a way that measure singularity issues are able to be circumvented. We will hence refer to this method as semi-centred.
The data is generated using a uniform mesh of $2^{18}$ points. We use the Karhunen-Loeve white noise representation, exploiting the explicit eigenstructure for this problem, and sample  $2^{16}$ coefficients and return the spatial field on a uniform mesh of the same size. We generate $2\times 10^5$ samples, with the first $5\times 10^4$ discarded as burn-in when calculating means. For the non-hierarchical method the value $\tau = 60$ is fixed, and for both hierarchical methods $\tau$ is initialised at this value. The level-set transformation of the posterior mean of the continuous field, spatially discretised as discussed above, are shown in \Cref{fig:gwf_truth}(top). That arising from the non-hierarchical method provides a fairly inaccurate reconstruction, due to the fixed length scale not matching that of the field to be recovered. Those arising from the hierarchical methods are similar to one another, both providing accurate reconstructions having learned appropriate length scales.
In \Cref{fig:gwf_truth}(bottom) the trace of $\tau$ is shown for both the semi- and non-centred hierarchical methods. The semi-centred chain takes approximately $5\times 10^4$ steps before $\tau$ reaches the region of high posterior probability, whereas the non-centred chain takes approximately 200 steps. Additionally, $\tau$ mixes worse with the semi-centred method. Thus, even though both methods are dimension-robust, the non-centred method provides much better statistical properties for  approximately the same computational cost. \Cref{fig:gwf_truth}(bottom) also illustrates how well the parameter $\tau$ is informed by the data, comparing the relatively flat prior density to the much more concentrated posterior density.

\begin{figure}[htbp]
\begin{center}
  \includegraphics[width=\textwidth, trim=2cm 0cm 2cm 0cm,clip]{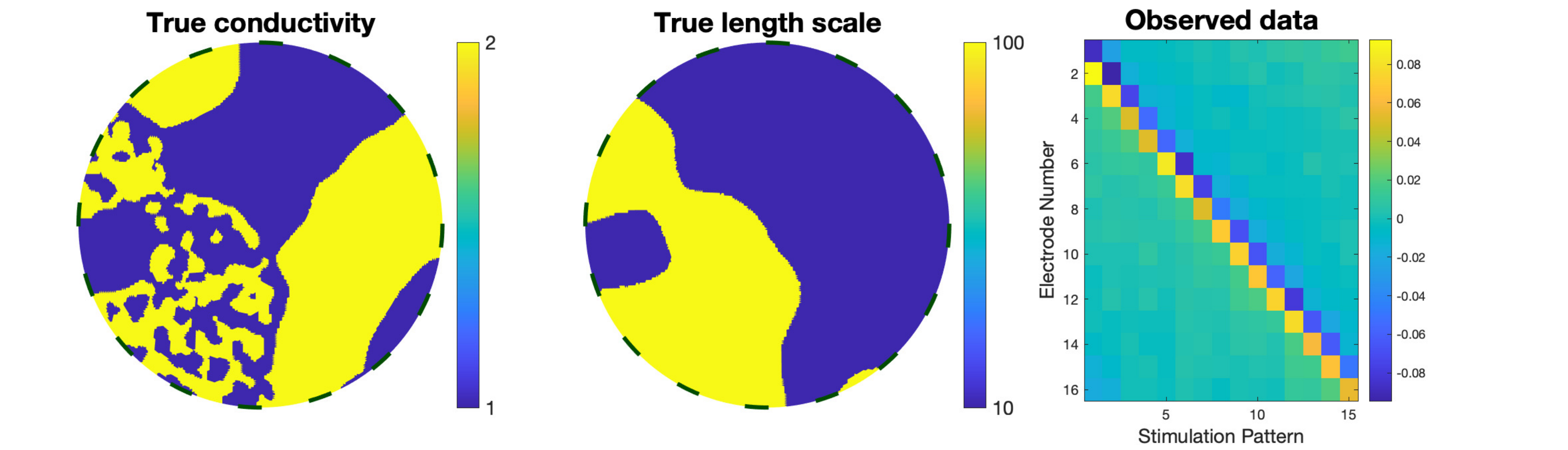}\\
\includegraphics[width=0.7\textwidth, trim=2cm 0cm 2cm 0cm,clip]{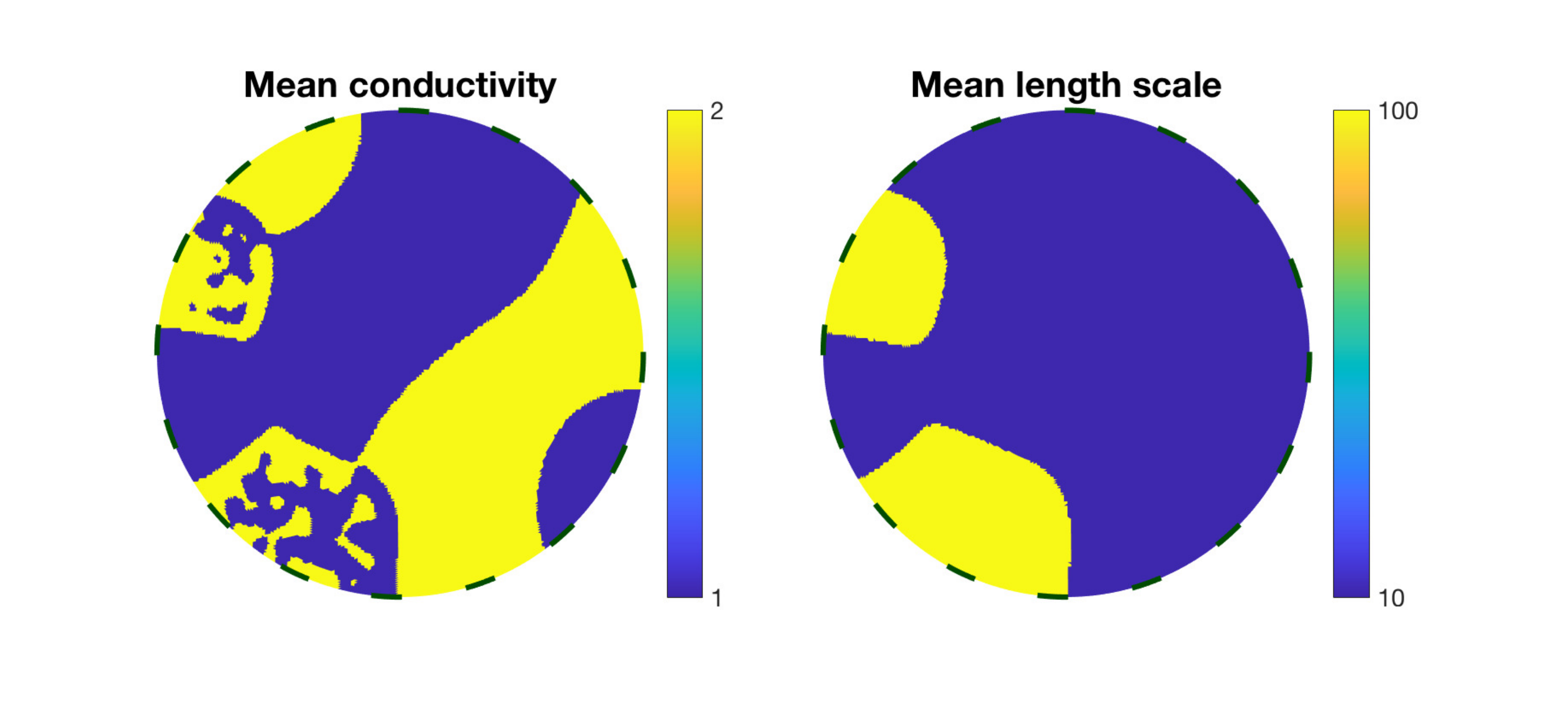}
\end{center}
\caption{Top: The true conductivity field (left), its associated length-scale field (middle), and the noisy data generated from this (right). Bottom: The level-set transformation of the posterior mean of the continuous conductivity field (left) and length-scale field (right).} 
\label{fig:eit_truth}
\end{figure}

We now showcase the proposed methodology on an example from medical imaging. 
We consider the Electrical Impedance Tomography (EIT) problem of recovering the interior conductivity of a body from voltage measurements on its boundary. Mathematically the model  is similar to \Cref{ex:inv_pde}, except measurements are made at the boundary for a variety of different boundary conditions, rather than in the interior. Full details of the model are given in \cite{somersalo1992existence}, and details of the Bayesian approach are provided in \cite{DS16a}. We focus on the task of recovering a binary conductivity with two distinct length scales associated with it. The observations are comprised of voltage measurements on each of the 16 electrodes for 15 different (linearly independent) current stimulation patterns; the top-right subfigure of \Cref{fig:eit_truth} shows these values concatenated into a matrix of size $16\times 15$.

If we were to use a thresholded hierarchical Whittle-Mat\'ern distribution as the prior, as in the permeability example, of the article, then the assertion of a constant length-scale throughout the body would restrict reconstruction accuracy. For our prior we hence assume in addition that the prior on the length-scale is itself a thresholded Whittle-Mat\'ern distribution with a fixed length-scale; effectively we have a ``deep'' Gaussian process prior.  
Such anisotropic length-scales make sense from the SPDE characterisation \cref{eq:matern_spde} of the distributions. Such priors have been considered without thresholding in \cite{roininen2016hyperpriors}.  \Cref{alg:mwg} assumes existence of a Lebesgue density of the hyperparameter. This is not the case here, though the methodology readily extends; see \cite{DGST17} for an explicit statement of the algorithm used. 
We take the true conductivity to be a draw from the prior, so that we may also examine the ability to recover the length-scale field. Observations are perturbed by white Gaussian noise with standard deviation $0.002$, leading to a large median relative error of approximately $21\%$. In \Cref{fig:eit_truth}(top) we show the true conductivity field we wish to recover, its length-scale field, and the observed data.
Samples are generated on the square $(-1,1)^2$ and restricted to the domain $D = B_1(0)$; this allows for some boundary effects to be ameliorated --  see \cite{roininen2014whittle,daon2016mitigating} for further discussion regarding boundary effects of this class of Gaussian random fields. Each field is sampled on a uniform mesh of $2^8\times 2^8$ points so that there are $2^{17}$ unknowns in total. The forward model is evaluated with a finite element method using the EIDORS software \cite{adler2006uses} on a mesh of $46656$ elements, with spline interpolation used to move from the sampling mesh.

We generate $4\times 10^5$ samples using the non-centred pCN within Gibbs method. The wpCN method may be used to update both fields simultaneously, however we found in practice that a Metropolis-within-Gibbs method led to better mixing. The level-set transformation  of the posterior  mean  is shown in \Cref{fig:eit_truth}(bottom); the first $1\times 10^5$ samples are discarded as burn-in in the calculation of these. The main features of the conductivity are recovered, and some information about the shape of the length-scale field is evidently contained within the data. The recovery of both fields is more accurate near to the boundary than the centre of the domain, as is to be expected from the nature of the measurements.


\section{Graph-based semi-supervised classification}
\label{sec:MNIST}

\subsection{Overview}

This Section illustrates the full potential of the computational methodology we introduce in this article and the underlying modelling framework. The primary objective is multi-class classification of $N=10200$ instances with as little supervision as possible; we illustrate our approach on the MNIST dataset, which is available from machine learning repositories and consists of thousands of  hand-written digits.  The data are in the form of input data, which are $20\times 20$ greyscale pixel intensities in $\{0,\ldots,255\}$, and output data, which are labels from 0 to 9; 20  of the available instances are shown in \Cref{fig:mnist_digits}.  In terms of modelling we employ a hierarchical level-set prior, as detailed in \Cref{sec:mnist_ip} below. In our formulation the latent states are $N$-dimensional, where $N=10200$; this is an example where the latent state is finite-dimensional but of high dimension. In terms of learning, we adopt a semi-supervised learning approach according to which we use labels for a tiny fraction of the available images and build probabilistic predictions for the rest. Therefore, although in this example all available data have been manually labelled, we use the dataset as testground for the more realistic and common situation where obtaining the labels is costly. We use an active learning approach to choose which data to obtain labels for by exploiting the uncertainty quantification aspect of our modelling/inference/computational approach: we first randomly sample 200 instances to label, then estimate labels and associated uncertainties for the remaining instances and then obtain labels for the instances that are associated with largest uncertainties.

\subsection{Inverse problem formulation}
\label{sec:mnist_ip}

We turn the input data into features by rescaling the pixel intensities so that each input observation is  a vector in $[0,1]^{400}$; then, as in \cite{bertozzi2017uncertainty} we project the datapoints onto their first $d = 50$ principal components; hence each input instance $x_j$ is a vector in $[0,1]^{50}$. The goal is to predict labels for $N=10200$ instances. With each class $r=1,\ldots,10$ we associate a latent state, $v_r$ which is an $N$-dimensional vector; therefore for each instance in the sample we associate 10 latent states that capture information about the instance's label. The mapping from latent states to labels is done using a vector level-set transformation, discussed in \Cref{sssec:lvl_prior}, 
\[
(Sv)(x) = \mathds{1}_{r(x;v)},\quad r(x;v) = \underset{r=1,\ldots,10}{\mathrm{arg}\max}\;v_r(x),
\]
where $\mathds{1}_r$ is a 10-dimensional vector with 0's except for the $r$th position in which it has an 1,  i.e.,  $\mathds{1}_r$ is the one-hot-encoding or the $r$th class. When available, the true class of the $j$th instance, $y_j$ will also be encoded in the same way, hence $y_j$ is also a 10-dimensional vector with nine 0's and one 1.

Our prior distribution on each $v_r$ is chosen to be Gaussian with covariance matrix that is constructed using input-data information.  We construct a graph with $N$ vertices $\{x_j\}_{j=1}^N$, and edge weights $\{w_{ij}\}_{i,j=1}^N$ given by some similarity function $w_{ij} = \eta(x_i,x_j)$ of the datapoints. Specifically let the weights be defined as the self-tuning weights introduced in \cite{zelnik2005self}. Define $L \in \R^{N\times N}$ to be the symmetric normalised graph Laplacian associated with this graph, i.e. $L = I - D^{-\frac{1}{2}}WD^{-\frac{1}{2}}$ where $W_{ij} = w_{ij}$ and $D = \mathrm{diag}_i(\sum_{j=1}^N w_{ij})$. The eigenvectors and eigenvalues of this matrix contain a priori clustering information about the input data; see \cite{von2007tutorial}. 

Let $\{\lambda_j\}_{j=0}^N$ denote the (non-negative) eigenvalues of $L$, ordered in a increasing way, and denote $\{q_j\}_{j=0}^N$ its corresponding eigenvectors. Given $\alpha > 0$ and $M \in \N$ with $M\leq N$, the random variable
\[
\sum_{j=0}^M \frac{1}{(1+\lambda_j)^{\alpha/2}}\xi_j q_j,\quad \xi_j \sim \sN(0,1)\text{ i.i.d.}
\]
has law $\sN(0,C(\alpha,M))$, where $C(\alpha,M) = P_M(I + L)^{-\alpha}P_M^*$, and $P_M$ is the matrix of top $M$ eigenvectors. The sum is analogous to the Karhunen-Lo\`eve expansion and defines directly the white noise representation of a random variable with this Gaussian distribution. We use this as a prior distribution for each $v_r$ of the unknown vector field, and take the $v_r$'s as a priori i.i.d..
We treat the parameters $\alpha,M$ hierarchically. A length-scale parameter could also be introduced and treated hierarchically, however we found empirically that this led to a lower classification accuracy in the mean. A non-hierarchical Bayesian approach to semi-supervised learning with this dataset was considered in \cite{bertozzi2017uncertainty}. There the prior covariance was fixed as $C = L^{-1}$, working on the orthogonal complement of $q_0$ so that the inverse is well-defined, and binary classification was considered on subsets of images of two digits. We shift $L$ by the identity to provide invertibility on the entirety of $\R^N$, increasing the flexibility of the prior.

We place uniform priors $\sU(1,100)$ and $\sU(\{1,2,\ldots,100\})$  on the hyperparameters $\alpha$ and $M$ respectively. The choice of prior on $\alpha$ is made to be generally uninformative, and that on $M$ is such that $M\ll N$. If the input data were perfectly clustered, which means that the associated graph defined earlier had disconnected components, the first few eigenvectors would suffice for perfect classification even with as little as a couple of dozen labelled instances, provided within cluster instances had the same label with high probability. This ideal is never the case in practice, hence we wish to consider enough eigenvectors.

We use the Bayesian level-set method with likelihood given by
\[
\Phi(v;y) = \frac{1}{2\gamma}\sum_{j=1}^J |(Sv)(x_j)-y_j|,
\]
with $\gamma = 10^{-4}$.
This is not the vector probit likelihood, rather it is a cheap approximation.  However, it has been shown in \cite{dunlop2018large} that the resulting posteriors from both choices converge weakly to the conditioned measure
\[
\mu(\dee v) \propto \mathds{1}(\{(Sv)(x_j) = y_j\,\,\forall j \in Z'\})\,\mu_0(\dee v)
\]
in the limit $\gamma\to 0$, and for the choice $\gamma = 10^{-4}$ inference with any of them leads to similar results. We start by using labels for a random sample of 200 instances. 

\begin{figure}
\begin{center}
\includegraphics[width=0.6\textwidth]{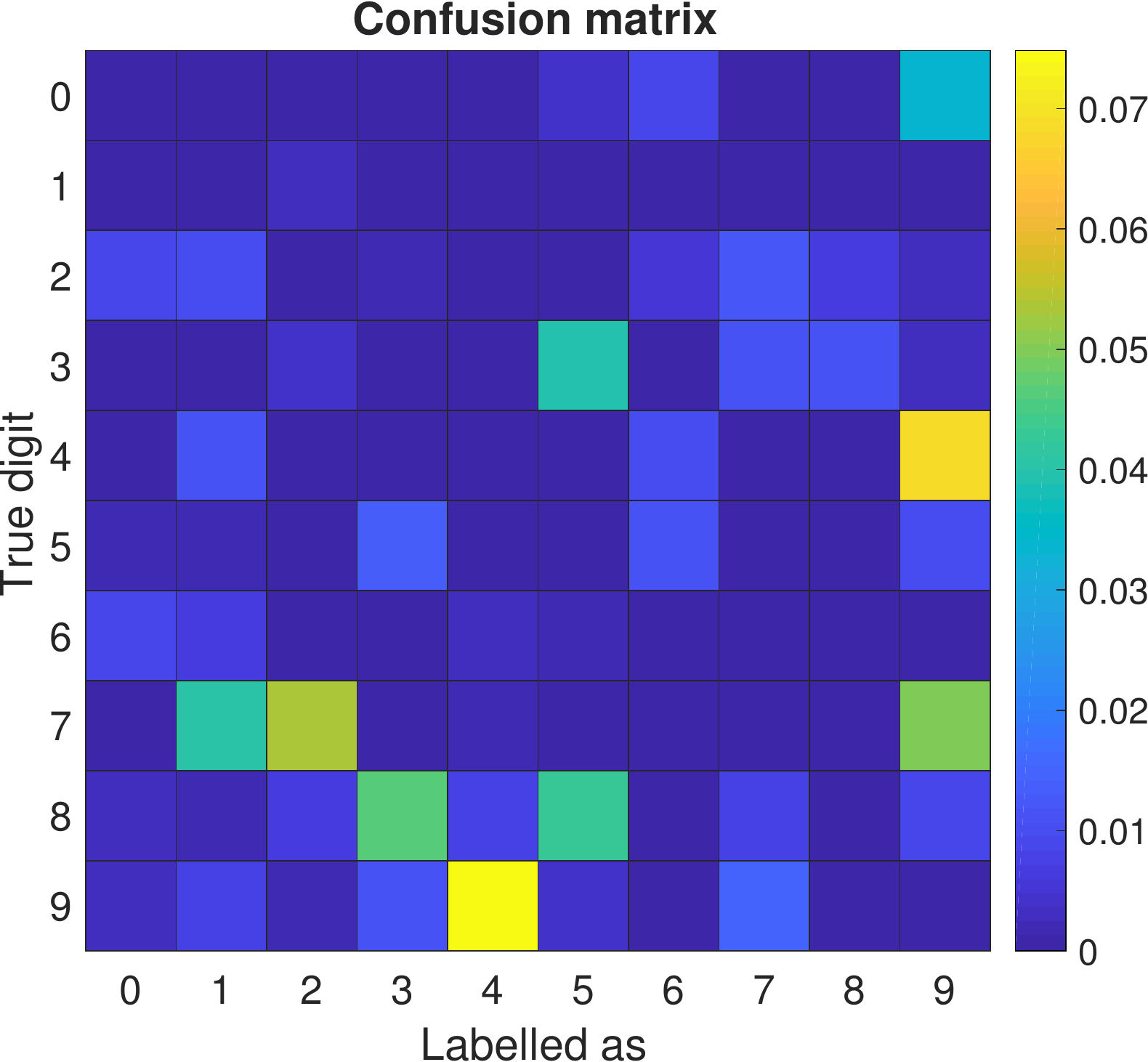}
\end{center}
\caption{The confusion matrix arising from the MNIST simulations}
\label{fig:mnist_confusion}
\end{figure}

We apply \Cref{alg:mwg}  using random walk proposals for the parameters and pCN for the latent states, to generate 70000 samples of which we discard 20000 samples as burn-in.  In \Cref{fig:mnist_confusion} we show a classification confusion matrix; the $(i,j)$th entry of this represents the percentage of images of digit $i$ that has been classified as digit $j$; we set the diagonal to zero to emphasise contrast between the off-diagonal entries. The most misclassified pairs are those digits that can look most similar to one another, such as 4 and 9, or 3 and 8.

\begin{figure}
\begin{center}
\includegraphics[width=\textwidth,trim=3.3cm 3.7cm 3.3cm 3.7cm, clip]{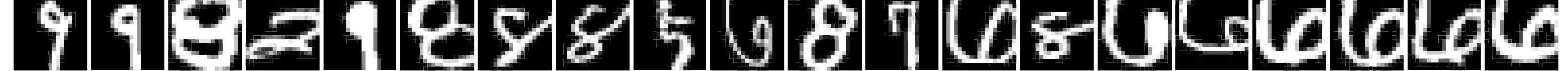}\vspace{0.2cm}
\includegraphics[width=\textwidth,trim=3.3cm 3.7cm 3.3cm 3.7cm, clip]{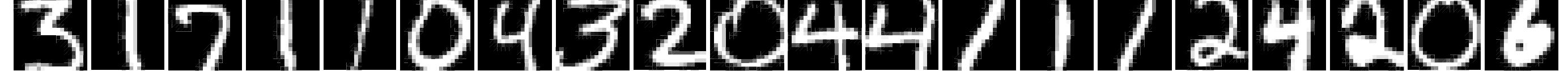}
\end{center}
\caption{The 20 most uncertain images (top) and a selection of 20 of the most certain images (bottom) under the posterior after labelling 200 images.}
\label{fig:mnist_digits}
\end{figure}

\begin{figure}
\begin{center}
\includegraphics[width=\textwidth,trim = 4cm 0cm 4cm 0cm,clip]{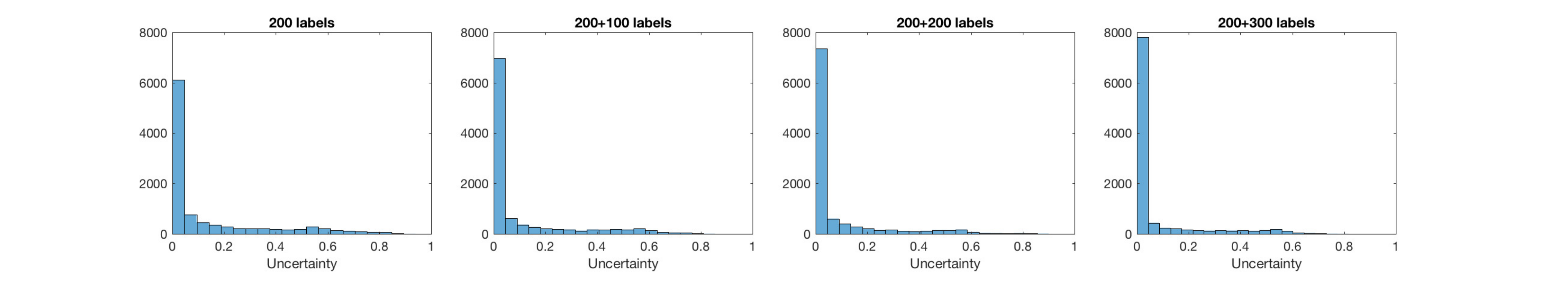}
\includegraphics[width=\textwidth,trim = 4cm 0cm 4cm 0cm,clip]{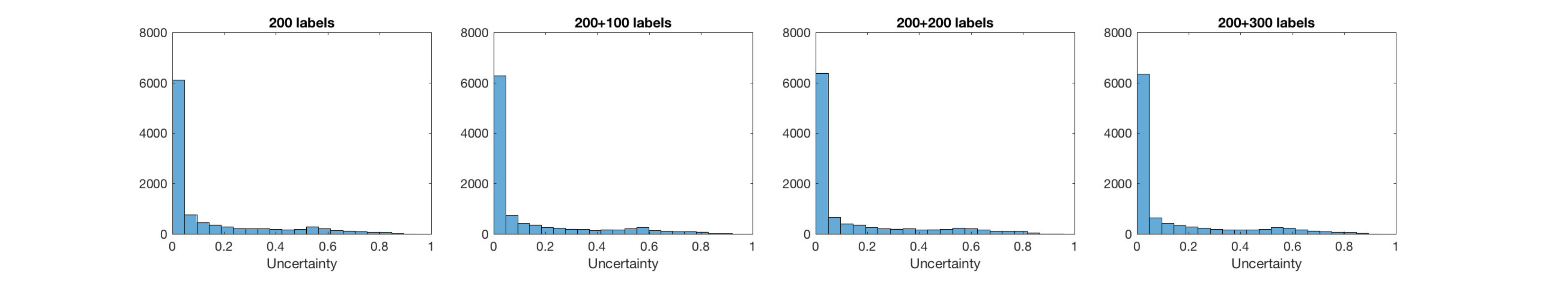}
\end{center}
\caption{Histograms illustrating the distribution of uncertainty across images as the number of labelled datapoints is increased using human-in-the-loop learning, iteratively labelling the most uncertain points (top) and most certain points (bottom).}
\label{fig:mnist_hist}
\end{figure}

We introduce the measure of uncertainty associated with a data point $x_j$ as
\[
U(x_j) = 1 - \frac{10}{9} \big\|\mathbb{E}(Sv)(x_j) - c\big\|_2^2
\]
where $c = \frac{1}{10}\sum_{r=1}^{10} \mathds{1}_r = (1/10,\ldots,1/10)$ is the centre of the simplex spanned by the classes. This is analogous to the variance of the classification in the case of binary classification as used in \cite{bertozzi2017uncertainty}.  The normalising factor ensures $U(x_j) \in [0,1]$ for all $x_j$.
The mean value of the uncertainty across all 10200 images is $0.135$. In \Cref{fig:mnist_digits} we show the 20 images with the highest uncertainty value, as well as a selection of 20 images with zero uncertainty value.  On the whole, the certain images appear clearer and more `standard' than the uncertain images, as would be expected. The uncertain images depict digits that have properties such as being sloped, cut off, or visually similar to different digits.
We can use this uncertainty measure to select an additional subset of the images to label, in an effort to decrease overall uncertainty in the classification. We now provide labels for the 100 most uncertain images, in addition to the original 200 images, and perform the MCMC simulations again. This could be interpreted as a form of human-in-the-loop learning, wherein an expert provides labels for data points deemed most uncertain by the algorithm, an idea that has been introduced in \cite{bertozzi2018uncertainty}.  This reduces the mean uncertainty across all images to $0.100$. As a comparison, we also consider labeling an additional set of the 100 most certain points rather than most the uncertain points, that is, points with $U(x_j) = 0$; this has a lesser impact, reducing the mean uncertainty to $0.128$.
In more detail, 
in \Cref{fig:mnist_hist} we show how the distribution of uncertainty changes when we perform the human-in-the-loop process iteratively, both for the most uncertain and most certain images. When the most uncertain images are labelled, the number of images in the bin of lowest uncertainty increases significantly. In particular, the first labelling of 100 additional images increases its size by over 1000. After labelling 300 additional images, the mean uncertainty has reduced to $0.070$. On the other hand, we see that labelling the most certain images does not change the distribution of uncertainty as much, and the mean uncertainty has become $0.132$ after labelling an additional 300 images.

\section{Conclusions}
\label{sec:conclusions}

The aim of this paper has been threefold. First, to introduce a plug-and-play MCMC sampling framework for posterior inference in Bayesian inverse problems with non-Gaussian priors. The user needs to specify the elements of the model and a white noise transformation. The proposed framework marries two important ideas in MCMC: non-centred parameterisations of hierarchical models and Metropolis-Hastings algorithms defined on infinite-dimensional spaces, such as the preconditioned Crank-Nicolson sampler. Second, to demonstrate the success of the approach in challenging problems where we showcase the desired robustness with respect to the dimension of the latent process, the advantages of the method over alternative plug-and-play methods, such as random walk Metropolis algorithms, and its comparable performance to tailored methods in models where such are available. Third, to showcase the wide range of applications that this methodology is appropriate for  that encompasses mainstream applications in Statistics, Applied Mathematics and Machine Learning.  
The sampling of the latent high-dimensional latent state relies on a single tuning parameter. The performance of the algorithms is relatively robust to its choice, provided the resultant acceptance probabilities are bounded away from 0 or 1. We make the case that for dimension-robust algorithms it is hard, maybe even infeasible, to obtain optimality theory for the choice of this tuning parameter, essentially for the same reasons why such a theory does not exist for low-dimensional sampling problems. The article also tackles a more challenging application where the full machinery of our methodology is employed. We cast multi-class semi-supervised classification as a Bayesian inverse problem, and use a Gaussian hierarchical prior for the latent variables that is informed by geometric features of the input data. We use the proposed sampling algorithms to obtain uncertainty quantification on the labelling of unlabelled data, which we then use to guide the choice of additional data to manually label in an active learning fashion. We believe that this application is characteristic of the potential of the cross-fertilization between scientific disciplines this article promotes.  

\noindent{\bf Acknowledgements} 
MMD and AMS are supported by AFOSR Grant FA9550-17-1-0185 and ONR Grant N00014-17-1-2079.

\bibliographystyle{plain}
\bibliography{bib_nc}

\section*{Appendix: Constructions for white noise, transformations and gradients}

This section provides a concise description of a function space setting which 
allows us to interpret the derivative of $\Psi(\cdot)$ required for the 
(conceptual) implementation of \Cref{alg:wmala} to sample elements
of a function space. Recall that $\Psi(\xi)=\Phi(T(\xi))$ and that $\xi$
is a white noise. In order to accomodate the white noise, we introduce the scale
of separable Hilbert-spaces $\cH^\tau$, parameterized by $\tau \in \R$, and given by
\[
\cH^\tau = \bigg\{u\in \R^\infty: \sum_j j^{2\tau} u_j^2 < \infty\bigg\}
\] 
with inner product $\la u,z\ra_{\cH^\tau} := \sum_j j^{2\tau} u_j z_j$.
Note that $\cH^0=\ell^2.$ 
White noise may be viewed as an infinite vector of i.i.d. standard Gaussians, $\xi_j\iid \sN(0,1)$ and hence does not live in $\ell^2$, almost surely. On the other hand
the identity covariance operator is trace class in $\cH^\tau$ for any  $\tau=-s$ with
$s>1/2$; in the following we fix such an $s$ and then draws from the measure 
are in $\cH^{-s}$ almost surely. 

We view $T: \cH^{-s} \to X$ as a bounded linear operator and
$\Phi$ as a nonlinear operator mapping $X$ into $\R$. The composition 
$\Psi=\Phi \circ T$ is then also a nonlinear operator mapping $\cH^{-s}$ into $\R$. Given a space $Z$, we denote $\cL(Z,Y)$ the space of linear functions from $Z$ to $Y$ and $\cL(Z,\R)$ the dual space of $Z$. If $Z=\cH^{-s}$ then its dual with respect to $\ell^2=\cH^0$ is 
(or more properly may be identified with) the Hilbert space $\cH^s.$ 

Let $\Psi'(\xi)$ denote the Gateaux derivative of $\Psi$ computed at $\xi$. For each fixed $\xi$ this is a linear operator in $\cL(\cH^{-s},\R)$, and so may be represented by an element $D\Psi(\xi) \in \cH^{s}$ via the dual pairing above; we refer to this representative as the gradient of $\Psi$ at $\xi$. In \Cref{alg:wmala}
the norm and inner-product should be interpreted as the norm in $\ell^2$ and the
dual pairing between $\cH^{-s}$ and $\cH^s$ respectively. These are both well-defined
since since  $D \Psi (\xi) \in \cH^s$. 

Note that in practice one will be provided with the derivative of $\Phi$ rather than directly with that of $\Psi$, however $D\Psi(\xi)$ may be computed from the derivatives of $\Phi$ and $T$ via the chain rule: $\Psi'(\xi) = \Phi'(T(\xi))\circ T'(\xi)$ with $\Phi'(T(\xi)) \in \cL(X,\R)$ and $T'(\xi) \in \cL(\cH^{-s},X)$. In terms of gradients, this may be written as $D\Psi(\xi) = T'(\xi)^*D\Phi(T(\xi))$, where we have again used the duality pairing to define the adjoint operator $T'(\xi)^*$ so that $D\Psi(\xi) \in \cH^{s}$. 

As an example, consider the case $T(\xi)= \sum_j \sqrt{\lambda_j} \xi_j \varphi_j$. This is linear and hence $T'(\xi)$ is independent of $\xi$ and  $T'(\xi)h  = \sum_j  \sqrt{\lambda_j} h_j \varphi_j$. The adjoint operator $T'(\xi)^*:X\to\cH^s$ may then be computed as $(T'(\xi)^*u)_j = \sqrt{\lambda_j} \la u,\varphi_j\ra_X$ for each $j$, so that
\[
D\Psi(\xi)_j = \sqrt{\lambda_j} \la D\Phi(T(\xi)),\varphi_j\ra_X.
\]
In the more general case $T(\xi) = m + \sum_j \rho_j \Lambda_j(\xi_j) \varphi_j$, we obtain \[
D\Psi(\xi)_j = \rho_j \Lambda_j'(\xi_j)\la D\Phi(T(\xi)),\varphi_j\ra_X.
\]

\end{document}